\documentclass[12pt]{article}

\RequirePackage{a4}
\usepackage{epsfig}
\usepackage[utf8x]{inputenc}
\usepackage{url}
\usepackage{here}
\usepackage{amssymb}
\usepackage{graphicx}
\usepackage{comment}
\usepackage{verbatim}
\usepackage{fancyvrb}
\usepackage{relsize}
\usepackage{makecell}

\usepackage{fancyhdr}

\usepackage{lineno}
\usepackage{hyperref}
\PassOptionsToPackage{hyphens}{url}\usepackage{hyperref}
\usepackage{breakurl}
\usepackage{xurl}
\usepackage{etoolbox}
\gappto{\UrlBreaks}{\UrlOrds}

\usepackage{amsmath}   
\usepackage{mathtools}

\VerbatimFootnotes

\begin{document}

\makeatletter
\def\maketitle{%
\cleardoublepage 
\thispagestyle{empty}%
\begingroup 
\topskip\z@skip
\null\vfil
\begin{center}
\LARGE\centering
\openup\medskipamount
\vspace{-38pt}\@title\par\vspace{15pt}%
\mdseries\large\@author\par \vspace{15pt}%
\vfill
\end{center}}
\makeatother

\title{\vspace{-1.2cm} 
  What is the probability that a vaccinated person is
  shielded from Covid-19?
  \\[0.2em]\smaller{}A Bayesian MCMC based reanalysis of published data\\
  with emphasis on what should be  reported as `efficacy'}
  
\author{Giulio D'Agostini\footnote{
    Universit\`a ``La Sapienza'' and INFN, Roma, Italia,
    giulio.dagostini@roma1.infn.it
  }
 \  and Alfredo Esposito\footnote{
   Retired, alfespo@yahoo.it
 }
}

\date{}

\maketitle

\begin{abstract}
Based on the information communicated in press releases, and finally
published towards the end of 2020 by Pfizer, Moderna and
AstraZeneca, we have built up a
simple Bayesian model, in which the main quantity 
of interest plays the role of {\em vaccine efficacy} (`$\epsilon$').
The resulting Bayesian Network is processed
by a Markov Chain Monte Carlo (MCMC), implemented in JAGS interfaced to
R via rjags. As outcome, we get several 
probability density functions (pdf's)
of $\epsilon$, each conditioned on the data provided by the three
pharma companies. The result is rather stable against large variations
of the number of people participating in the trials and it is
`somehow' in good agreement with the results provided by the companies,
in the sense that their values correspond to the
most probable value (`mode') of the pdf's resulting from MCMC,
thus reassuring us about the
validity of our simple model. 
However we maintain that the number to be reported as {\em vaccine efficacy}
should be the mean of the distribution, rather than the mode,
as it was already very clear to Laplace about 250 years ago
(its `rule of succession' follows from the simplest problem of the kind).
This is particularly important in the case in which the number
of successes equals the numbers of trials, as it happens with
the  efficacy
against `severe forms' of infection, claimed by
Moderna to be 100\%.
The implication of the various
uncertainties on the predicted number of vaccinated infectees
is also shown, using both MCMC and approximated formulae. 
\end{abstract}
{\small 
 \mbox{}\mbox{}\\
 \rightline{{\em ``\ldots the most important questions of life
     \ldots }\hspace{1 cm}}\\
 \rightline{{\em are indeed for the most part only problems
     in probability''}\hspace{1 cm}}\\
\mbox{}\\
\rightline{{\em ``we find that an event having occurred
    successively}\hspace{1 cm}}\\
\rightline{{\em any number of times, the probability that it
    will happen again}\hspace{1 cm}}\\
\rightline{{\em the next time is equal to this number increased
    by unity}\hspace{1 cm}}\\
\rightline{{\em divided by the same number, increased
    by two units''}\hspace{1 cm}}\\
\rightline{ (Laplace)\hspace{1 cm}}
}
\setcounter{footnote}{0}

\section{Introduction}
Our perspectives about living with Covid-19 changed dramatically
in the last months of 2020 with the results from the vaccine trials 
by Pfizer and Moderna and, a bit later, by AstraZeneca.
The first one claimed a
90\% efficacy \cite{Pfizer_09nov} (then updated to 95\% in
a further press release \cite{Pfizer_18nov} and in a much more
detailed paper \cite{Pfizer_NEJM}) and 
the second one 94.5\% \cite{Moderna_16nov} 
(later updated to 94.1\% \cite{Moderna_30nov, Moderna_NEJM}), 
while AstraZeneca press released and then published
two values (90.0\% and 62.1\%) depending on two
(unplanned \cite{AZ1, AstraZeneca}) different experimental
settings.\footnote{In one of the experimental setting,
  indicated here as `low dose'-`standard dose', the
  first vaccine dose was half of the planned one
  (`standard dose'-`standard dose' setting).}

In an unpublished first paper\footnote{The paper
is available on the web site
of one of the authors, together with the slides of a related webinar
and the code to reproduce the results~\cite{vaccini}.} based on the
first press releases by
Pfizer and Moderna, we remarked that, since the announcements did
not mention any uncertainty,  we
understood that the initial Pfizer's number was the result of a rounding,
with uncertainty of the order of the percent. But then, we continued,
we were highly surprised by the Moderna's announcement,
providing the tenths of the percent, as if it were much more precise.
We had indeed the impression that the `point five'
was taken very seriously, not only by media speakers, who put the
emphasis on the third digit, but also by experts
from which we would have expected a phrasing implying
some uncertainty in the result (see e.g. Ref.~\cite{Fauci_yt_16nov}).
The same remarks apply to the later AstraZeneca announcement.
In fact, a fast exercise shows that, in order to have an uncertainty
of the order of a few tenths of percent, the number of
vaccine-treated individuals that got the Covid-19 had to be
at least of the order of {\em several hundreds}.
But this was not the case. In fact, the actual numbers were indeed
much smaller, 
as we learned from the Moderna first press release~\cite{Moderna_16nov}:
``\textsl{This first interim analysis was based on 95 cases,
of which 90 cases of COVID-19 were observed in the placebo
group versus 5 cases observed in the mRNA-1273 group, resulting
in a point estimate of vaccine efficacy of 94.5\%
($p <0.0001$)}''.

Now, it is a matter of fact that if a physicist reads for
an experimental result 
a number like `5', she tends to associate to it, as a rule of thumb,
an uncertainty of the order of its square root, that is $\approx 2.2$.
Applied to the Moderna claims, this implies an {\em inefficacy} of
{\em about} $(5.5\pm 2.3)\%$, or an efficacy of
{\em about} $(94.5\pm 2.3)\%$.
Another reason that made us worry about the result was, 
besides the absence of an  associated uncertainty~\cite{ISO},
the ``$p <0.0001$'' accompanying it,
being us extremely critical against
{\em p-values} and other frequentist prescriptions
(see Ref.~\cite{WavesSigmas} and references therein).

\begin{table}[t]
\mbox{}   
\begin{center}
  \begin{tabular}{|l|c|c|r|r|c|c|}
 \hline   
 & \makecell{release\\date} &\makecell{sample size\\vaccine - placebo}&$n_{V_I}$ &  $n_{P_I}$ & $n_{V_{I_s}}$ &  $n_{P_{I_s}}$ \\
 \hline 
  Moderna-1 \cite{Moderna_16nov}&Nov 16& 14134 - 14073  & 5  & 90   & -- & -- \\
  Moderna-2 \cite{Moderna_30nov, Moderna_NEJM}&Nov 30& {\em (same)}& 11 & 185  & 0  & 30 \\
  Pfizer \cite{Pfizer_18nov, Pfizer_NEJM}&Nov 18  & 18198 - 18325   &  8  &  162 & 1 & 9 \\
  AstraZeneca LD-SD \cite{AZ1, AstraZeneca} &Nov 23& 1367 -  1374 & 3  & 30  & -- & -- \\
  AstraZeneca SD-SD \cite{AZ1, AstraZeneca} &Nov 23& 4440 -  4455 & 27 & 71   & -- & -- \\
 \hline 
\end{tabular}    
\end{center}
\caption{\small \textsf{Bare data concerning the number of
  infected in the vaccine group ($n_{V_I}$) and  in the placebo group ($n_{P_I}$).
  `Moderna-1' is just an {\em interim result}, based on the same
  sample of `Moderna-2', that
  we had used in Ref.~\cite{vaccini}.
  In the case of Moderna and Pfizer comprehensive results,
  also the numbers for the occurrence of `severe forms of infection'
  were reported ($n_{V_{I_s}}$ and $n_{P_{I_s}}$, respectively). In the case
  of AstraZeneca LD-SD stands for `low dose followed by standard dose',
  SD-SD for `two consecutive standard doses'.} 
  }
  \label{tab:data}
\end{table}
\begin{table}[t]
\mbox{}   
\begin{center}
  \begin{tabular}{|l|c|c|}
 \hline   
 & efficacy value  & 95\% `uncertainty interval'   \\
 \hline 
  Moderna-1 \cite{Moderna_16nov} & $ 0.945$ & ---------- \\
 Moderna-2 \cite{Moderna_NEJM} & $ 0.941$ & $[0.893, 0.968]$
 ({\em confidence} interval) \\
 Pfizer \cite{Pfizer_NEJM} &  $0.950$ & $[0.903, 0.976]$
 \  ({\em credible} interval)\ \ \  \\
  AstraZeneca LDSD \cite{AZ1} & $0.900$ & $[0.674, 0.970]$
   ({\em confidence} interval)\\
   AstraZeneca SDSD \cite{AZ1}& $0.621$ & $[0.410, 0.757]$
    ({\em confidence} interval)\\
 \hline 
\end{tabular}    
\end{center}
\caption{\small \sf Published vaccine efficacy
  (value and 95\% `uncertainty interval'). }
  \label{tab:vaccine_efficacy_published}
\end{table}

In our first paper \cite{vaccini} we tried then to understand whether
it was possible to get an idea
of the {\em possible values of efficacy} consistent with the 
data, each one associated with its degree of belief on the basis of
the few data available in those days.
In other words, our purpose was and is to arrive to
a probability density function (pdf), although not obtained
in closed form, of the quantity of interest. 

In the present paper we not only extend our analysis
 to the published data%
~\cite{Pfizer_NEJM, Moderna_NEJM, AstraZeneca}
(see Tab.~\ref{tab:data}), but can also  compare
our results with the published ones, summarized in 
Tab.~\ref{tab:vaccine_efficacy_published},
which also include an indication of the
uncertainty to be associated with them. 
What makes us confident about the validity
of our simple model is that the press released and finally
published results concerning `efficacy'
(see Tab.~\ref{tab:vaccine_efficacy_published})
are in excellent agreement
with the mode of the distribution we
get analyzing our model through a
Markov Chain Monte Carlo (MCMC).
This is not a surprise to us, indeed.
In fact we are aware of statistical methods
which tend to produce as `estimate' the most probable value
of the quantity of interest, that would be inferred starting
from a flat prior~\cite{BR}. The fact that different kind
of `uncertainty intervals' are provided will be discussed at the due
point. We only anticipate here that they have {\em in this case}  
equivalent meaning.

The paper is organized as follows. In Sec.~\ref{sec:model}
we describe and show how to implement in JAGS~\cite{JAGS}
the causal model connecting in a probabilistic way the
quantities of interest, among which the primary role is played
by the `efficacy' $\epsilon$. We also give, in footnote \ref{fn:ContoEsatto},
some indications
on how to proceed in order to get exact results for $f(\epsilon)$,
although they can only be obtained numerically. 
The MCMC results are 
shown and discussed in Sec.~\ref{sec:MCMCresults}.
Then the question asked in the title is tackled,
with a didactic touch and including some
historical remarks, in Sec.~\ref{sec:mode_mean}. 
The observation that the resulting pdf's of $\epsilon$
can by approximated rather well by Beta distributions
(Sec.~\ref{sec:BetaApprox}) leads us to discuss in further detail
the role of the {\em priors}, initially chosen simply uniform.
Then Sec.~\ref{sec:expected_NVI} is devoted
to the related question of predicting the number of vaccinated people
that shall result infected, taking into account several sources of uncertainty. 
Finally, in Sec.~\ref{sec:severity} we extend our analysis
to the level of protection
given by the vaccines against the disease severity,
in which the outcome of a simple application of probability theory is at odds 
with simplistic, extraordinary claims.\footnote{ \,`{\em All 30
    cases occurred in the placebo group and
    none in the mRNA-1273 vaccinated group.}'
  `{\em ...and vaccine efficacy against severe
    COVID-19 was 100\%}' (Moderna press release~\cite{Moderna_30nov}, 
    based on no severe cases out of the 11 infectees in the vaccine group).
This result has been reported
as 100\% efficacy with (uncritical!) great emphasis also in
the media~\cite{Moderna_100_media} -- a reminder of the C.~Sagan's
quote that
``{\em extraordinary claims require extraordinary evidence}''
is here in order.\label{fn:extraordinary}}
In Sec.~\ref{sec:Summing} we sum up the analysis strategy and
the outcome of the paper, also commenting on the optimal sharing
of vaccine/placebo samples in the test trial. 
Then some conclusions and final remarks follow.

\section{Model and analysis method}\label{sec:model}
Dealing with problems of this kind,
we have learned (see e.g. Ref.~\cite{sampling})
the importance of  building up a graphical representation
of the {\em causal model} relating the quantities
of interest, some of them `observed' and others `unobserved',
among the latter the quantities we wish to infer.
Also in this case, despite some initial skepticism
about the possibility of getting some meaningful results, 
once we have built up the model, very basic indeed,
it was clear that the main outcome concerning the vaccine efficacy
was not depending on the many aspects of the trials.
Our initial doubts were in fact related to the several
details concerning
the people involved in the test campaign,
but they finally resulted
to be much less critical than we had at first thought.

The causal model used in this analysis
is implemented in the  {\em Bayesian network}
of Fig.~\ref{fig:BayesianNet}. 
\begin{figure}  
  \begin{center}
    \epsfig{file=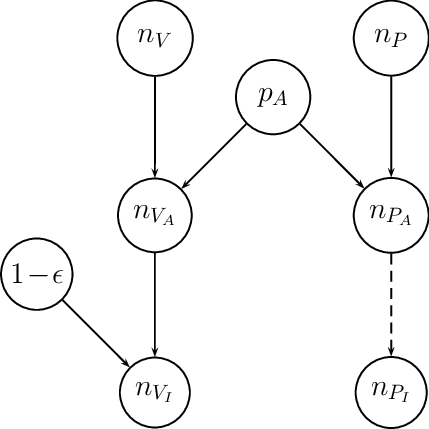,clip=,width=0.45\linewidth}
      \\  \mbox{} \vspace{-1.0cm} \mbox{}
  \end{center}
  \caption{\small \sf Simplified Bayesian network of the vaccine vs placebo
    experiment (see text). 
  }
  \label{fig:BayesianNet}
\end{figure}
The top {\em nodes} $n_V$ and $n_P$ stand for the
number of individuals in the vaccine and placebo (i.e. control)
groups, respectively,
as the subscripts indicate, while the bottom
ones ($n_{V_I}$ and $n_{P_I}$) are the number of individuals
of the two groups 
resulting infected during the trial. These are the observed nodes
of our model, whose values are summarized in Tab.\ref{tab:data}.

Then, there is the question of how to relate the numbers
of infectees to the numbers of the participants in the trial. 
This depends in fact on several variables, like the 
prevalence of the virus in the population(s) of the involved people,
their social behavior, personal 
life-style, age, health state and so on. And, hopefully,
it depends on the fact
that a person has been vaccinated or not.
Lacking detailed information, we simplify
the model introducing an {\em assault probability} $p_A$, that is a  
catch-all term embedding the many real life variables, apart
being vaccinated or not. 
Nodes $n_{V_A}$ and $n_{P_A}$
in the network of Fig.~\ref{fig:BayesianNet}
represent then the number of `assaulted individuals'
in each group, and they are modeled according to
binomial distributions, that is 
\begin{eqnarray}
  n_{V_A} &\sim & \mbox{Binom}(n_V, p_A) \label{eq:nVA_da_NV}\\
  n_{P_A} &\sim & \mbox{Binom}(n_P, p_A)\,,   \label{eq:nPA_da_NP}
\end{eqnarray}
represented in the graphical model by solid arrows.

The `assaulted individuals' of the control group
are then assumed to be all infected, and hence the
deterministic link with dashed arrow relating  node $n_{P_A}$
to node $n_{P_I}$ follows (indeed the two numbers are exactly
the same in our model, and we make this distinction
only for graphical symmetry with respect to the vaccine group).

Instead, the  `assaulted individuals' of the other group
are `shielded' by the vaccine with probability $\epsilon$,
that we therefore identify with {\em efficacy},
although we shall come back at the due point about what 
should be reported as `efficacy'.
The probability of becoming
infected {\em if assaulted} is therefore equal to $1\!-\!\epsilon$,
so that node $n_{V_I}$ is related to node $n_{V_A}$ by
\begin{eqnarray}
  n_{V_I} &\sim & \mbox{Binom}(n_{V_A}, 1\!-\!\epsilon). \label{eq:nVI_da_NVA }
\end{eqnarray}    
At this point all the rest is a matter of calculations,
that we do by MCMC techniques\footnote{
Indeed one could try to
get an exact solution for the pdf of $\epsilon$.
The steps needed are: write down the joint pdf of all
the variables in the network; condition on the certain variables;
marginalize over all the uncertain variables besides $\epsilon$.
Referring to Ref.~\cite{sampling} for details, here is
the structure of the unnormalized pdf obtained starting
from uniform priors over $\epsilon$ and $p_A$:
\begin{eqnarray*}
f(\epsilon\,|\,n_V,n_P,n_{V_I},n_{P_I}) &\propto&
\sum_{n_{V_A}=n_{V_I}}^{n_V}\int_0^1\!\mbox{d}p_A
\left[\frac{n_{V_A}!}{(n_{V_A}-n_{V_I})!}\cdot(1-\epsilon)^{n_{V_I}}\cdot
  \epsilon^{n_{V_A}-n_{V_I}}\right] \cdot \\
&& \mbox{}\hspace{2.05cm}
\left[\frac{1}{n_{V_A}!\,(n_V-n_{V_A})!}\cdot
  p_A^{n_{V_A}}\cdot (1-p_A)^{n_V-n_{V_A}}\right] \cdot \\
&& \mbox{}\hspace{2.05cm}\left[p_A^{n_{P_A}}\cdot (1-p_A)^{n_P-n_{P_A}}\right]
\,,
\end{eqnarray*}
where the three terms within square brackets are the
three binomial distributions entering the model, stripped
of all irrelevant constant factors. Simplifying and reorganizing the 
various terms we get 
\begin{eqnarray*}
f(\epsilon\,|\,n_V,n_P,n_{V_I},n_{P_I}) &\propto&
(1-\epsilon)^{n_{V_I}}\cdot
\sum_{n_{V_A}=n_{V_I}}^{n_V}\frac{\epsilon^{n_{V_A}-n_{V_I}}}
                        {({n_{V_A} -n_{V_I})}!\,(n_V-n_{V_A})!}\cdot \\
&&\int_0^1\mbox{}
p_A^{n_{V_A}+{n_{P_A}}}\cdot (1-p_A)^{n_V-n_{V_A}+{n_P-n_{P_A}}}\,\mbox{d}p_A\,.
\end{eqnarray*}
We recognize that the integral
$\int_0^1x^{r-1}\cdot (1-x)^{s-1}\,\mbox{d}x$,
in terms of a generic variable $x$,
defines the special function {\em beta}\,
$\mbox{\small \tt B}(r,s)$, thus obtaining  
\begin{eqnarray*}
f(\epsilon\,|\,\ldots) \!\!&\propto&\!\!
(1-\epsilon)^{n_{V_I}}\cdot\!\!\!
\sum_{n_{V_A}=n_{V_I}}^{n_V}\frac{\epsilon^{n_{V_A}\!-n_{V_I}}}
                        {({n_{V_A}\! -n_{V_I})}!\,(n_V\!-n_{V_A})!} \cdot 
                        \mbox{\small \tt B}(n_{V_A}\!\!+\!{n_{P_A}}\!\!+\!1,\,
                        n_V\!\!-\!n_{V_A}\!+n_P\!-\!n_{P_A}\!\!+\!1).
\end{eqnarray*}
Then the integral over $\epsilon$
follows, in order to get the normalization factor.
Finally, all moments of interest can be evaluated.
All this can be done numerically. However, we proceed to MCMC, being its use much
simpler and also for the flexibility it offers
(for example in the case we need to extend the model,
as we shall do in Secs.~\ref{sec:expected_NVI} and \ref{sec:severity}).
\label{fn:ContoEsatto}
}
with the help of the program  JAGS~\cite{JAGS}
interfaced with R~\cite{R} via rjags~\cite{rjags}.

The nice thing using such a tool is that we have to take
care only to describe the model, with
instructions whose meaning is quite
transparent:\footnote{Those who have no experience with
JAGS can find in Ref.~\cite{sampling} several ready-to-run R scripts.} 
\begin{verbatim}
      model {
        nP.I  ~ dbin(pA, nP)           # 1.          
        nV.A  ~ dbin(pA, nV)           # 2.
        pA    ~ dbeta(1,1)             # 3. 
        nV.I  ~ dbin(ffe, nV.A)        # 4.  [ ffe = 1 - eff ]
        ffe   ~ dbeta(1,1)             # 5.
        eff   <- 1 - ffe               # 6. 
      }
\end{verbatim}
We easily recognize in lines 1. and 2. of the R code
the above Eqs.~(\ref{eq:nVA_da_NV}) and
(\ref {eq:nPA_da_NP}), while line 4. stands for
Eq.~(\ref{eq:nVI_da_NVA }). Line 6. is simply the transformation
of `$1\!-\!\epsilon$' (`{\rm ffe}' in the code) to $\epsilon$,
the quantity we want to trace in the `chain'. 
Finally lines 3. and 5. describe the {\em priors} of the
`unobserved nodes' that have no `parents', in this case
$p_A$ and $1\!-\!\epsilon$.
We use for both a {\em uniform prior},
modeled by a \mbox{Beta} distribution (see Sec.~\ref{ss:BayesProblem}
for details) with
parameters $\{1,\,1\}$.\footnote{We cannot go here into the details
of this choice that we consider
quite reasonable, given the information provided by
the data, and refer for the details to Ref.~\cite{sampling}
and references therein. The fact that, as we shall
  see in next section, the modes of the distributions of $\epsilon$
  that result from our analysis practically coincide
  with the efficacy values reported by the three companies
  means that they have also used `flat priors', or
  frequentist methods which {\em implicitly entail}
  a flat prior~\cite{BR}.
  We shall see in Sec.~\ref{ss:informative_priors} how `informative priors'
  (e.g. by experts) can be taken into account in a second step,
  without the need of repeating the analysis for each choice of priors.
  \label{fn:flat_priors}
} 
Then we have to provide the data, in our case
$n_V$, $n_P$,  $n_{V_I}$ and  $n_{P_I}$.
The program samples the space of possibilities
and returns lists of numbers (a `chain')
for each `monitored variable', which can then
be analyzed `statistically'. For example 
the frequency of occurrence of the values
in each list is expected to be proportional
to the probability of that values of the variable
(Bernoulli's theorem). Similarly we can evaluate
correlations among variables.

\begin{table}[!t]
  \begin{center}
  MCMC results
  \vspace{0.2cm}\mbox{}
  \begin{tabular}{|l|c|c|c|}
 \hline   
 & \makecell{mean $\pm$  `stand.\,unc.'} & centr. 95\% `cred.\,int.' &
 $P(\epsilon \ge 0.9)$  \\
 \hline 
  {Moderna-1} & $ 0.933 \pm 0.028 $ & $[0.866, 0.976]$ & 0.875 \\
  Moderna-2 & $ 0.935 \pm 0.019 $ & $[0.892, 0.967]$ & 0.951 \\
  Pfizer &  $0.944 \pm 0.019$ & $[0.900, 0.975]$ & 0.974\\
  AstraZeneca LDSD & $0.861 \pm 0.075$ & $[0.678, 0.964]$ & 0.349\\
  AstraZeneca SDSD & $0.599 \pm 0.090$ & $[0.400, 0.750]$ & 0.000\\
 \hline 
  \end{tabular}
  \mbox{}\vspace{0.8 cm}\mbox{}
  Published results
  \vspace{0.2cm}\mbox{}
   \begin{tabular}{|l|c|c|}
 \hline   
 & efficacy value  & 95\% `uncertainty interval'   \\
 \hline 
  Moderna-1 \cite{Moderna_16nov} & $ 0.945$ & ---------- \\
 Moderna-2 \cite{Moderna_NEJM} & $ 0.941$ & $[0.893, 0.968]$
 ({\em confidence} interval) \\
 Pfizer \cite{Pfizer_NEJM} &  $0.950$ & $[0.903, 0.976]$
 \  ({\em credible} interval)\ \ \  \\
  AstraZeneca LDSD \cite{AZ1} & $0.900$ & $[0.674, 0.970]$
   ({\em confidence} interval)\\
   AstraZeneca SDSD \cite{AZ1}& $0.621$ & $[0.410, 0.757]$
    ({\em confidence} interval)\\
 \hline 
   \end{tabular}
  \mbox{}\vspace{-0.2cm}\mbox{} 
\end{center}
\caption{\small \sf Top table: MCMC results for the 
model parameter $\epsilon$ (see text).  
 Bottom table: same as Tab.~\ref{tab:vaccine_efficacy_published} 
  for easier comparison with the MCMC results.
  }
  \label{tab:vaccine_efficacy}
\end{table}
\section{MCMC based results} \label{sec:MCMCresults}
We run the model on the basis of the published bare data 
summarized in Tab.~\ref{tab:data}.
The MCMC results concerning the `efficacy parameter' $\epsilon$ are
summarized in Tab.~\ref{tab:vaccine_efficacy}, to be compared with the published results, shown in Tab.~\ref{tab:vaccine_efficacy_published} and repeated below our ones for the reader's convenience.
 
The MCMC based pdf's of $\epsilon$ 
are plotted in Fig.~\ref{fig:vaccine_efficacy}
with smooth curves showing the profile of the histograms
of the $\epsilon$ values in the chains.
\begin{figure}[t]
  \begin{center}
    \epsfig{file=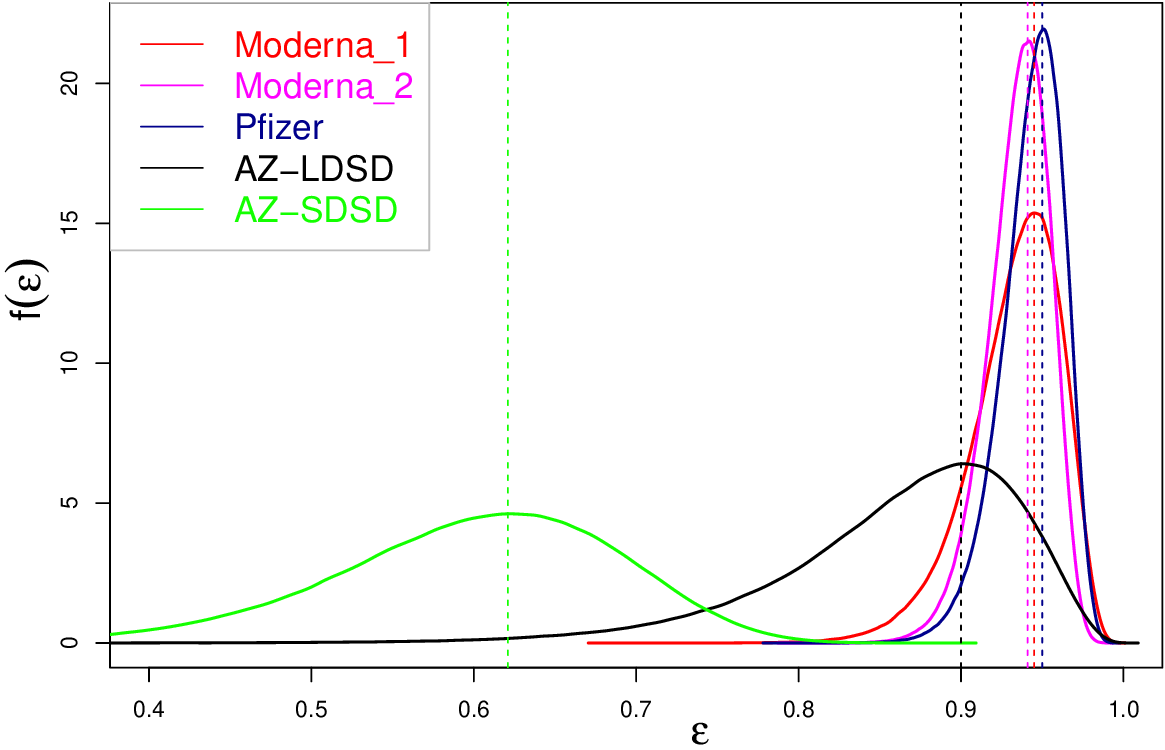,clip=,width=\linewidth}
      \\  \mbox{} \vspace{-1.2cm} \mbox{}
  \end{center}
  \caption{\small \sf MCMC results for the vaccine efficacies.
    The vertical lines indicate the results provided by the
    pharma companies, in practical perfect agreement with the
    mode of the MCMC estimated probability distributions. 
  }
  \label{fig:vaccine_efficacy}
\end{figure}
For comparison, the vertical dashed lines show the
results of the  pharma companies (`efficacy value'
in Tab.~\ref{tab:vaccine_efficacy_published}). 
As we can see, they  correspond practically
exactly to the {\em modal values} of the distributions. 
This makes us quite confident about the validity of our simple
model for this quantitative analysis, although 
we maintain that the single number for the efficacy to be provided
is not the mode, but rather the mean of the distribution,
as we shall argue in Sec.~\ref{sec:mode_mean}.
However, some remarks are in order already at this point. In fact,
although there is no doubt about the fact
that the most complete description of a probabilistic inference
is given by the pdf of the quantity of interest,
it is also well understood that it is often convenient
to summarize the distribution with just a few numbers.

Usually, when inferring physical quantities, 
the preference goes to the {\em mean}  and the
{\em standard deviation} (the latter being
related to the concept of {\em standard uncertainty}~\cite{ISO})
because of rather general
probability theory theorems
which make their use convenient for further evaluations
(`propagations', as we shall also see in Sec.~\ref{sec:expected_NVI}).
Other ways to summarize with just a couple of
numbers a probability distribution are intervals which contain
the uncertain value of the variable of interest at a given probability level
({\em credible interval}). We report then in
Tab.~\ref{tab:vaccine_efficacy}
the 95\% {\em central credible interval}\,\footnote{The
  meaning of such an interval is that,
  conditioned on the data used and on the model assumptions,
  we consider $P(\epsilon < \epsilon_{low})
  = P(\epsilon > \epsilon_{high}) = 2.5$\%,
  where $\epsilon_{low}$ and $\epsilon_{high}$ are the
  boundaries of the interval.
}
evaluated from the
MCMC chains as well as the 90\% `right side credible interval'.
Other useful summaries, depending on the problem of interest,
can be the most probable value of the distribution (mode) and the 
{\em median}, i.e. the value that divides the possible values
into two equally probable intervals.
As we have stated above, the modes of the MCMC based pdf's  
coincides with the values reported as `efficacy value' in
Tab.~\ref{tab:vaccine_efficacy_published}, which contains
also what we have  generically indicated as 95\% `uncertainty interval',
in form of credible interval for Pfizer
and {\em confidence interval} for the other
two companies.\footnote{It is important to understand that,
  strictly speaking, a 95\% frequentistic
  confidence interval does not provide the interval
  in which the authors are
  95\% {\em confident} that the `true value' of interest lies
  (see Refs.~\cite{WavesSigmas,BR} and references therein), although
  it is `often the case' for `routine measurements'~\cite{BR}. 
}

The MCMC  also  provides results for the other
`unobserved' nodes of the causal model, in our case $p_A$ and
$n_{V_A}$. We refrain from quoting results on the `assault probability',
because they could easily be misunderstood, as they strongly
depend, contrary to
$\epsilon$, on the values of $n_V$ and $n_P$,
being $p_A$ a catch-all quantity embedding several
real life variables, {\em including} the virus prevalence.
We have however checked that our main results on $\epsilon$ are stable against
the (simultaneous) variations of $n_V$ and $n_P$ by orders of magnitude
(thus implying similar large variations of $p_A$).\footnote{In other words,
  the gross result essentially depends on the ratios
$n_{V_I}$:$n_{P_I}$ and $n_{V}$:$n_{P}$.}

We give, instead,
the results concerning $n_{V_A}$ that we expect to be {\em around}
 $n_{P_I}$. We get, in fact, respectively for Moderna-1, Moderna-2, Pfizer,
AstraZeneca (LDSD) and AstraZeneca (SDSD) the following values:
$89\pm 13$, $185\pm 19$, $160\pm 18$, $29\pm 8$ and $70\pm 12$
(note that the standard uncertainty
is not simply the root square of $n_{V_I}$, as a rule of thumb would suggest).

\section{What is the probability that a vaccinated person gets
shielded from Covid-19?}\label{sec:mode_mean}

It is now time to come to the question asked in the title.
We have already used the noun
{\em efficacy},  associated to the uncertain variable
$\epsilon$ of our model of Fig.~\ref{fig:BayesianNet}.
Then, analyzing the published data, we have got by MCMC several
pdf's of $\epsilon$, that is $f(\epsilon\,|\,\mbox{Moderna-1})$,
$f(\epsilon\,|\,\mbox{Moderna-2})$, and so on
(see Fig.~\ref{fig:vaccine_efficacy}).
Hereafter, since what we are going to say is rather general,
we shall indicate the generic pdf by
$f(\epsilon\,|\,\mbox{data},H)$, where $H$ stands
for the set of {\em hypotheses}\footnote{In general we are used
  to indicating by
  $I$ the {\em background state of information}\,\cite{sampling},
  but we reserve here the symbol $I$ for `infected'.}
underlying our inference and not specified in detail.

Let us now focus on the probability that an {\em assaulted individual}
gets infected. Indicating by $A$ the
condition `the individual is assaulted', by $V$ the condition `vaccinated'
and by $I$ the event `the individual gets infected'
(and therefore $\overline{A}$,  $\overline{V}$ and  $\overline{I}$
their logical negations), we get, rather trivially,
\begin{eqnarray}
P(I\,|\,\overline{A}) &=& 0\,,
\end{eqnarray}
while, in the case of assault, the probability of infection
depends on whether the individual has been vaccinated or not.
In the case of placebo, following our model, we  simply get
\begin{eqnarray}
  P(I\,|\,A,\overline{V}) &=& 1\,. 
\end{eqnarray}
Instead, in case the individual has been vaccinated,
the probability of infection will depend on $\epsilon$, that is,
for the special cases of {\em perfect shielding} and
{\em no shielding} (i.e. no better than the placebo),
\begin{eqnarray}
  P(I\,|\,A,V,\epsilon=1) &=& 0   \label{eq:eff_0} \\
  P(I\,|\,A,V,\epsilon=0) &=& 1\,.  
\end{eqnarray}
In general, if we {\em were} certain about the {\em precise value}
of $\epsilon$, the probability of getting infected or
not is related in a simple way to $\epsilon$:
\begin{eqnarray}
  P(I\,|\,A,V,\epsilon) &=& 1-\epsilon \\
  P(\overline{I}\,|\,A,V,\epsilon) &=& \epsilon\,.  \label{eq:eff_eff}
\end{eqnarray}
The above equations, and in particular
Eq.~(\ref{eq:eff_eff}), express in mathematical
terms the meaning we associate to {\em efficacy}, in terms
of the {\em model parameter} $\epsilon$: {\em the probability
  that a vaccinated person gets shielded from a virus}
(or from any other agent).
But the value of $\epsilon$
cannot be known precisely. It is, instead, affected by an uncertainty,
as it (practically) always happens for results of
measurements~\cite{ISO} (and indeed also 
the pharma companies accompany their results with uncertainties
-- see Tab.\ref{tab:vaccine_efficacy_published}).
In a probabilistic approach, this means that there are values of $\epsilon$
we believe more and values we believe less. All this, we repeat it,
is summarized by the probability density function
$$f(\epsilon\,|\,\mbox{data},H)\,.$$
The way to take into account all possible values of
$\epsilon$, each weighted by $f(\epsilon\,|\,\mbox{data},H)$,
is to follow the rules of probability theory, i.e.
\begin{eqnarray}
  P(\overline{I}\,|\,A,V,\mbox{data},H) & = &
  \int_0^1\!P(\overline{I}\,|\,A,V,\epsilon)\cdot f(\epsilon\,|\,\mbox{data},H)\,\mbox{d}\epsilon\,. \nonumber
\end{eqnarray}
Using then Eq.~(\ref{eq:eff_eff}) we get 
\begin{eqnarray}
  P(\overline{I}\,|\,A,V,\mbox{data},H) 
   & = &\int_0^1\!\epsilon\cdot f(\epsilon\,|\,\mbox{data},H)\,\mbox{d}\epsilon\,,
  \label{eq:integral_over_epsilon}
\end{eqnarray}
which represents {\em the probability
  that a vaccinated person}, \underline{not} belonging to the
trial sample, {\em gets shielded from Covid-19}, on the basis of
the data obtained from the trial and all (possibly reasonable)
hypotheses assumed in the data analysis.
It is easy to understand that $P(\overline{I}\,|\,A,V,\mbox{data},H)$
is what really
matters and therefore what should be communicated as
{\em efficacy} to the scientific community
and to the general public.\footnote{In this paper we only focus
  on efficacy, without even trying to enter
  on the related topics of {\em effectiveness},
  that refers to how well the vaccine performs
  in the real world (see e.g. Ref.~\cite{Gavi}),
  that is influenced by several other factors.}

Now, technically, 
Eq.~(\ref{eq:integral_over_epsilon}) is nothing but the {\em mean}
of the distribution of $\epsilon$. This should then be 
the number to report, and \underline{not} the mode
of the distribution, which has no immediate probabilistic meaning
for the questions of interest.\footnote{Let us repeat once more
that the result of frequentistic point estimates can be easily shown
to be equivalent, under reasonable assumptions, to the mode of the distribution
obtained by a probabilistic analysis.}

Now, if we compare the `efficacy values' of
Tab.~\ref{tab:vaccine_efficacy_published} with the mean values
of Tab.~\ref{tab:vaccine_efficacy} we see that in most cases
the differences are rather small (about 1/3 to 1/2 of
a standard deviation),
although the modal values
(to which, as we have showed above, the published efficacies correspond)
are always a bit higher than the mean values, due to the
left skewness of the pdf's. Therefore our point is mostly 
methodological,\footnote{However, we wish to point out that, from the practical
point of view, what really matters is the {\em inefficacy} 
i.e. the probability of getting infected
(see Sec. \ref{sec:expected_NVI} for details).  
Indeed, even though two
hypothetical values of $\epsilon$ equal to $0.96$ and $0.98$,
respectively, might appear quite close, nevertheless they imply
that the relative probabilities of getting infected are
in the ratio 2:1.} with some worries
when the mean value and the most probable
one differ significantly.

\subsection{An old {\em Problem in the Doctrine of Chances}}
\label{ss:BayesProblem}
Evaluating the probability of future events on the basis
of the outcomes of previous trials on `apparently the same conditions'
is an old, classical problem
in probability theory that goes back to about 250 years ago
and it is associated to the names of Bayes~\cite{Bayes} and
Laplace~\cite{Laplace_PC}. The problem can be sketched
as considering events whose probability of occurrence depends
on a parameter which we generically indicate as $p$, i.e.
\begin{eqnarray*}
  P(E\,|\,p) &=& p\,.
\end{eqnarray*}
Idealized examples of the kind are the
proportion of white balls in a box containing a large number
of white and black balls (with the extracted ball
put back into the box after each extraction),
the bias of a coin and the ratio of the chosen
surface in which a ball thrown `at random' can stop, 
with respect to the total surface of a horizontal table
(this was the case of the Bayes' `billiard', although the
Reverend did not mention a billiard).

A related problem concerns the number of times
(`$X$') events of a given kind
occur in $n$ trials, assuming that $p$ remains constant. The result is given by the
well known binomial, that is
\begin{eqnarray}
  X &\sim & \mbox{Binom}(n, p)\,, \label{eq:BinomLaplace}
\end{eqnarray}
whose graphical causal model is shown in the left
diagram of Fig.~\ref{fig:binom_and_inv}.
\begin{figure}[t]
  \begin{center}
    \epsfig{file=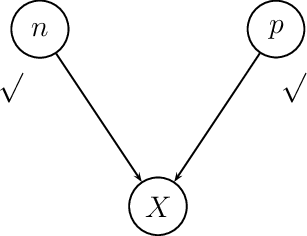,clip=,width=0.35\linewidth}
    \hspace{1.8cm}
    \epsfig{file=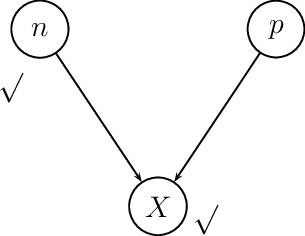,clip=,width=0.35\linewidth}
       \\  \mbox{} \vspace{-1.0cm} \mbox{}
  \end{center}  
  \caption{\small \sf Graphical models of the binomial
    distribution (left) and its `inverse problem'. The symbol
    `$\surd$' indicates the `observed' {\em nodes}
    of the {\em network}, that is the value of the quantity
    associated to it
    is ({\em assumed to be}) certain. The other node
    (only one in this simple case)
    is  `unobserved' and it is associated to
    a quantity whose value is uncertain.
  }
   \label{fig:binom_and_inv} 
\end{figure}

The problem first tackled in quantitative terms by Bayes and Laplace
was how to evaluate the probability of a `future' event $E_f$,
based on the information that in the past $n$ trials
the event of that kind occurred $X=x$ times (`number of successes')
and on the assumption of a {\em regular flow from past
  to future},\footnote{A reminder of Russell's {\em inductivist turkey}
  is a must at this point!} 
that is assuming $p$ constant although uncertain. In symbols,
we are interested in 
\begin{eqnarray*}
  P(E_f\,|\,n,x,H)\,,
\end{eqnarray*}
where $H$ stands, as above, for all underlying
hypotheses. Both Bayes and Laplace realized that
the problem goes through two steps: first finding the probability
distribution of $p$ and then evaluating $ P(E_f\,|\,n,x,H)$ taking into
account all possible values of $p$. In modern terms
\begin{eqnarray}
  1.\  \  \rightarrow \ \ \ \ f(p\,|\,n,x,H) &&  \label{eq:step_1} \\
  \ \ \ \ 2. \ \  \ \rightarrow  \ P(E_f\,|\,n,x,H)
        &=& \int_0^1 \!P(E_f\,|\,p)\cdot f(p\,|\,n,x,H) \label{eq:step_2}\\
        &=& \int_0^1 p \cdot f(p\,|\,n,x,H) \,.   \label{eq:step_2_a}
\end{eqnarray}
The basic reasoning behind these two steps
is expressly outlined in the {\em Sixth} and 
{\em Seventh Principle of the Calculus of Probabilities}, expounded by
Laplace in Chapter III of his
{\em Philosophical Essay on Probabilities}\,\cite{Laplace2}:
\begin{itemize}
\item the {\em Sixth Principle}, in terms of the possible causes $C_i$
  responsible of the observed event $E$, is essentially
  what is presently known as
  Bayes' theorem, that is
   \begin{eqnarray}
       P(C_i\,|\,E) &=& \frac{P(E\,|\,C_i)\cdot P_0(C_i)}
                 {\sum_k P(E\,|\,C_k)\cdot P_0(C_k)}\,,
      \label{eq:Principle6}
   \end{eqnarray}
   in which $P_0(C_i)$ is the so called {\em prior} probability of $C_i$,
   i.e. not taking into account the piece of {\em information}
   provided by the observation of $E$. Note
   that the role of $P_0(C_i)$ was explicitly considered by Laplace, who 1) before
   gave the rule in the case of $P_0(C_i)$ numerically all equal, which then drop
   from Eq.~(\ref{eq:Principle6}); 2) then specified that ``{\em if these various
     causes, considered {\sl à priori}, are unequally probable,
     it is necessary, in the place of the probability of the event resulting from
     each cause, to employ the product
     of this probability by the possibility of the cause itself}.''
   (Here `possibility' and `probability' are  clearly used as synonyms.) 
   Then, the importance of the finding is stressed:
   \begin{quote}
     ``{\em This is the fundamental principle of this branch of the analysis
       of chances which consists in passing from events to causes.}''
   \end{quote}
   Generalizing this `principle' to an infinite number of causes,
   associated to all possible values of the parameter $p$, with the
   `event' being the observation of $X=x$ successes in $n$ trials,
   we get the case sketched in the right
   diagram of Fig.~\ref{fig:binom_and_inv}, in which the unobserved
   node is now $p$. 
   Equation (\ref{eq:Principle6}) becomes then,
   in terms of the probability function of $X$ and of the
   pdf of $p$ $[$for which we take the freedom
     of using the same symbol `$f()$'$]$,
    \begin{eqnarray}
       f(p\,|\,n,x) &=& \frac{f(x\,|\,n,p)\cdot f_0(p)}
                 {\int_0^1 f(x\,|\,n,p)\cdot f_0(p) \,\mbox{d}p}\,.
      \label{eq:Principle6_cont}
   \end{eqnarray}
 \item The {\em Seventh Principle} then states that ``{\em the probability of
   a future event is the sum of the products of the probability of each cause,
   drawn from the event observed, by the probability that, this cause existing, the future
   event will occur}'', that is
    \begin{eqnarray}
       P(E_f) &=& \sum_i P(C_i)\cdot P(E_f\,|\,C_i)\,.
      \label{eq:Principle7}
    \end{eqnarray}
   Generalizing also this `principle' to an infinite number of causes
   associated to all possible values of the parameter $p$
   we get Eq.~(\ref{eq:step_2}), and then  Eq.~(\ref{eq:step_2_a}):
   {\em the probability of interest is the mean of the distribution of $p$}.
\end{itemize}
The solution of Eq.~(\ref{eq:Principle6_cont}), in the case
$X$ is described by Eq.~(\ref{eq:BinomLaplace}) and 
we consider 
all values of $p$ {\em à priori} equally likely, is a Beta pdf,
that is\footnote{See e.g. Ref.~\cite{sampling} and references therein.}
\begin{eqnarray}
       p &\sim & \mbox{Beta}(r,s)       \label{eq:Beta_p}
\end{eqnarray}
with $r=x+1$ and $s=n-x+1$.
Mean value 
and variance of the possible values of $p$ are then
\begin{eqnarray}
\mu \equiv \mbox{E}(p)&=&\frac{r}{r+s} \label{eq:Ebeta}\\
\sigma^2 \equiv \mbox{Var}(p)&=&\frac{r\cdot s}{(r+s+1)\cdot(r+s)^2}\,.
\label{eq:Varbeta}
\end{eqnarray}
Finally, using  Eq.~(\ref{eq:step_2}) and  Eq.~(\ref{eq:step_2_a})
we get the Laplace's {\em rule of succession}
\begin{eqnarray}
P(E_f\,|\,n,x,H) &=& \frac{x+1}{n+2}\,.\label{eq:succession}
\end{eqnarray}
Thus, in the special case of `$n$ successes in $n$ trials',
``{\em we find that an event having occurred successively any
  number of times, the probability that it will happen
  again the next time is equal to this number increased
  by unity divided by the same number, increased by two units}''~\cite{Laplace2},
i.e.
\begin{eqnarray}
P(E_f\,|\,n,x=n,H) &=& \frac{n+1}{n+2}\,.
\end{eqnarray}
In the case of $x=n=11$ we have then
12/13, or 92.3\%. Reporting thus 
100\% (see footnote \ref{fn:extraordinary})
can be at least misleading,
especially because such a  value can be (as it {\em has indeed been})
nowadays promptly broadcasted uncritically
by the media (see e.g. Ref.~\cite{Moderna_100_media} -- we have heard
so far no criticism in the media of such an incredible claim,
but only sarcastic comments by colleagues).

\section{Beta approximation of the MCMC results
and its utility}\label{sec:BetaApprox}
Moving to our results about the `model parameter $\epsilon$'
(it is now time to be more careful with names), reported
in  Tab.~\ref{tab:vaccine_efficacy} and Fig.~\ref{fig:vaccine_efficacy},
it should now be clear why the number to report as {\em efficacy}
should be the mean of the distribution. As far as the distribution
of $\epsilon$ is concerned, given the similarity of the inferential problem
that was first solved by Bayes and Laplace, we have good reasons
to expect that it should not `differ much' from
a Beta. In order to test the correctness of our guess we have done the
simple exercise of superimposing over the MCMC distributions
of Fig.~\ref{fig:vaccine_efficacy} the Beta pdf's evaluated
from mean and standard deviation of Tab.~\ref{tab:vaccine_efficacy}.
The distribution parameters can be
in fact obtained solving Eqs.~(\ref{eq:Ebeta})\,-\,(\ref{eq:Varbeta})
for $r$ and $s$:\footnote{To make it clear, no `fit' on the MCMC histogram
has been performed.}
\begin{eqnarray}
  r &=& \frac{(1-\mu)\cdot \mu^2}{\sigma^2} - \mu
  \label{eq:r_from_E-sigma}\\
  s  &=&  \frac{1-\mu}{\mu}\cdot \left[\frac{(1-\mu)\cdot \mu^2}{\sigma^2} - \mu\right]\,.
    \label{eq:s_from_E-sigma}
\end{eqnarray}
\begin{figure}[t]
  \begin{center}
    \epsfig{file=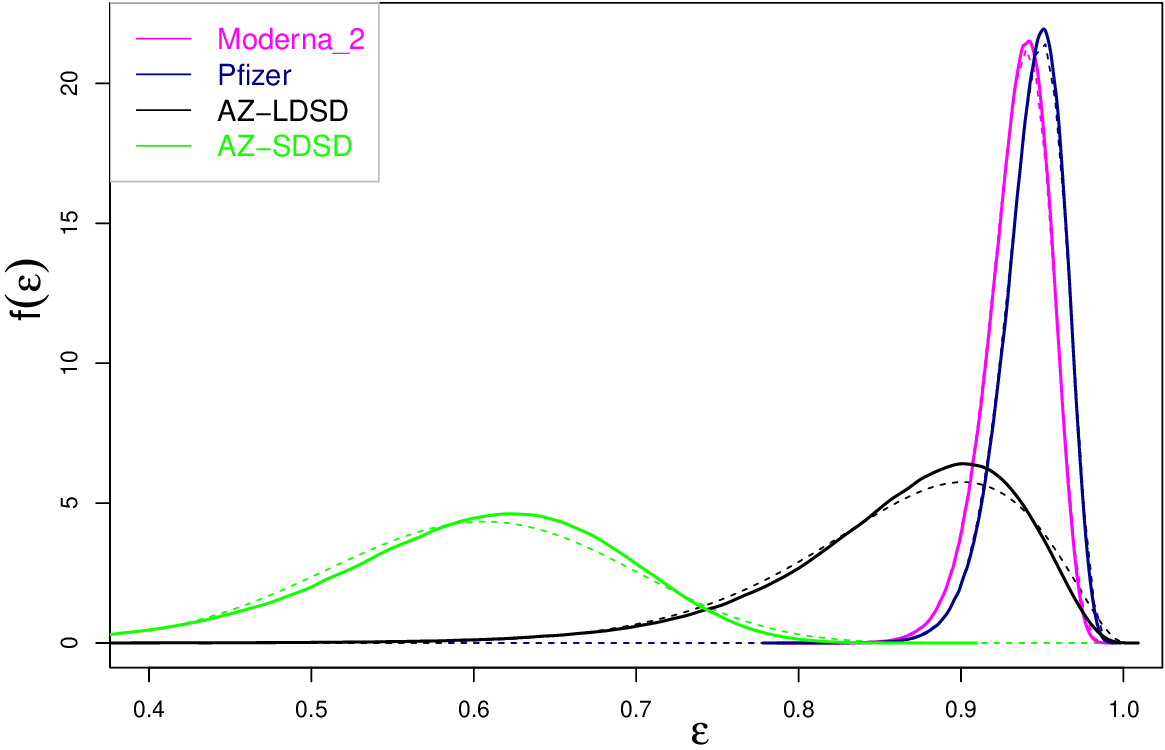,clip=,width=\linewidth}
    \end{center} 
    \caption{\small \sf MCMC inferred distributions of $\epsilon$
      (solid lines exactly as in Fig.~\ref{fig:vaccine_efficacy}) with superimposed
      (dashed lines, often coinciding with the solid ones)
      the corresponding Beta distributions evaluated from
      the mean values and the standard deviations resulting from MCMC.
    } \label{fig:epsilon_beta}
\end{figure}
The result is shown in Fig.~\ref{fig:epsilon_beta}.
As we can see, the agreement is  rather good for all cases,
especially for Moderna and Pfizer, for which the Beta and MCMC curves
practically coincide.

\subsection{Taking into account `informative priors' about $\epsilon$}
\label{ss:informative_priors}
The fact that the MCMC results are described with high degree
of accuracy by a Beta distribution
is not only a sterile curiosity, but has indeed
an interesting practical consequence.

As we have seen,
the pdf's of $\epsilon$ have been obtained starting from a uniform
prior. The same must be for Pfizer, since they also did a
Bayesian analysis, as explicitly stated in their paper
\cite{Pfizer_NEJM} and as revealed by the expression
`credible interval' (see Tab.~\ref{tab:vaccine_efficacy_published}),
and their values practically
coincide with ours. Instead, in the case of the other results, the expression
`confidence interval' seems to refer to a
frequentistic analysis, in which ``there are no priors''.
But in reality it is not difficult to show that sound
frequentist analyses (e.g. those based on {\em likelihood})
can be seen as approximations of Bayesian analyses
in which a flat prior was used (see e.g. Ref.~\cite{BR}).
The resulting `estimate' corresponds to the
mode of the posterior distribution under that assumption.

The question is now what to do if an expert has a `non flat'
{\em informative prior} (indeed none would  {\em à priori} believe that
values of $\epsilon$ close to zero or to unity would be equally likely!).
Should she ask to repeat the analysis inserting her
prior distribution of $\epsilon$?
Fortunately this is not the case. Indeed, as we have discussed
in Ref.~\cite{sampling}, due to the symmetric and peer roles
of {\em likelihood} and prior in the so called Bayes' rule, 
{\em each of the two has the role of `reshaping' the other}. 
Moreover, since a posterior distribution based on a uniform prior
concerning the variable of interest can be interpreted
as a likelihood (besides factors irrelevant for the inference),
we can {\em apply} to it an {\em expert's prior in a second time}
(see  Ref.~\cite{sampling} for details).
It becomes then clear the importance of
the observation that the pdf's of $\epsilon$
derived by MCMC can be approximated by Beta distributions:
from the MCMC  mean and standard deviation
we can evaluate the parameters of the
Beta of interest, as we have seen above;
this function can be then easily multiplied by the expert's prior;
the normalization can be done
by numerical integration and finally the posterior
distribution of $\epsilon$ also conditioned on the
expert's prior can be obtained.

This implementation in a second step
of the expert opinion becomes particularly simple if also 
her prior is modeled by a Beta, recognized to be a quite 
flexible distribution. For example, indicating
by $r_F$ and $s_F$ the Beta parameters $[$calculated with
Eq.~(\ref{eq:r_from_E-sigma})\,-\,(\ref{eq:s_from_E-sigma})$]$
obtained by a flat prior and by
$r_0$ and $s_0$ the parameters of the Beta
informative prior, the posterior distribution
will still be a Beta with parameters
\begin{eqnarray}
r_p &=& r_0 + r_F -1 \label{eq:update_rp}\\
s_p &=& s_0 + s_F -1\,, \label{eq:update_sp}
\end{eqnarray}
as it can be easily shown.\footnote{In fact, assuming
  that the MCMC based pdf of $\epsilon$ starting from a flat prior (`$F$')
  can be approximated by a Beta, we can write it,   
  neglecting irrelevant factors, as
\begin{eqnarray*}
f_F(\epsilon) &\propto& \epsilon^{r_F-1}\cdot (1-\epsilon)^{s_F-1}\,.
\end{eqnarray*}
Expressing also the informative prior by a Beta, that is
\begin{eqnarray*}
f_0(\epsilon) &\propto& \epsilon^{r_0-1}\cdot (1-\epsilon)^{s_0-1}\,,
\end{eqnarray*}
and applying the Bayes' rule, we get for the posterior (`$p$')
\begin{eqnarray*}
  f_p(\epsilon) &\propto& f_F(\epsilon) \times f_0(\epsilon) \\
  &\propto&  \epsilon^{r_F-1+r_0-1}\cdot (1-\epsilon)^{s_F-1+s_0-1} \\
   &\propto&  \epsilon^{(r_F+r_0-1)-1}\cdot (1-\epsilon)^{(s_F+s_0-1)-1}\,,
\end{eqnarray*}
from which Eqs.~(\ref{eq:update_rp})\,-\,(\ref{eq:update_sp}) follow.
}

\section{Expected numbers of vaccinated individuals
that shall become infected}\label{sec:expected_NVI}
Once we have got the efficacy of a vaccine, or, even better,
the full information about $\epsilon$ inferred by the
data, we can tackle another interesting problem: if we vaccinate $n_V'$
individuals in (possibly) another population, how many of them
will become infected?
  Obviously, this depends not only on $f(\epsilon)$
but also on many other parameters that we can model with the
{\em assault probability} $p_A'$ in the new population, and, obviously,
on the fact that $n_V'$ must be only a small part of the entire population,
so that we do not have to consider issues related  to herd immunity.
In order to do this exercise we need to enlarge
our causal model, thus getting the one shown  
in Fig.~\ref{fig:numero_atteso_vaccinati_infetti}.
\begin{figure}[t]
\begin{center}
    \epsfig{file=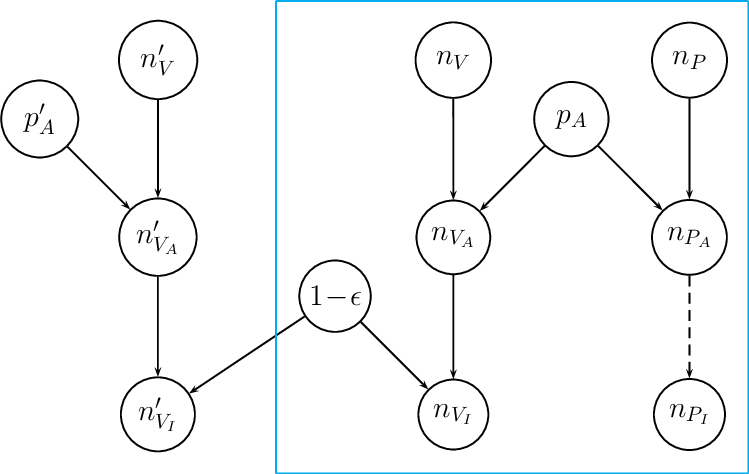,clip=,width=0.7\linewidth} \\
    \caption{\small \sf Variation of the model of Fig.~\ref{fig:BayesianNet},
      in this figure inside a box, 
      in order to consider the effect of the vaccine on $n_V'$ individuals
    of a different population (see text).}
    \label{fig:numero_atteso_vaccinati_infetti}
  \end{center}
\end{figure}

But, indeed, there is no need to set up a new JAGS model
and rerun the MCMC. We can just use the chain of $\epsilon$
obtained processing through the MCMC 
the original model of Fig.~\ref{fig:BayesianNet},
and do the remaining work with `direct' Monte Carlo.
However, having observed that the resulting pdf of $\epsilon$
is approximated a Beta, we can do it starting from
the mean and standard deviation obtained from the MCMC.
Here is e.g. how the evaluation of $n'_{V_I}$
can be performed by sampling (in the R code we
have indicated $n_V'$ as {\tt nV}, and so on): 
\begin{verbatim}
mu <- 0.944; sigma <- 0.019; e.rep <- 0.950    # Pfizer
# mu <- 0.935; sigma <- 0.019; e.rep <- 0.941    # Moderna
# mu <- 0.861; sigma <- 0.075; e.rep <- 0.900    # AZ LDSD
# mu <- 0.599; sigma <- 0.090; e.rep <- 0.621    # AZ SDSD

# uncomment the following line to simulate a negligible uncertainty 
# sigma <- 0.0001   

r = (1-mu)*mu^2/sigma^2 - mu
s  = r*(1-mu)/mu
cat(sprintf("r = %.2f, s =  %.2f\n", r, s))

ns <- 1000000
nV <- 100000  
pA <- 0.01

nA <- rbinom(ns, nV, pA)
cat(sprintf("nA: mean+-sigma:  %.1f +- %.1f\n", mean(nA), sd(nA)))

eps <- rbeta(ns,r,s)
nvI <- rbinom(ns, nA, 1-eps)
hist(nvI, nc=100, col='cyan', freq=FALSE, main='')
cat(sprintf("nvI: mean+-sigma:  %.1f +- %.1f\n", mean(nvI), sd(nvI)))
lines(rep(pA*nV*(1-mu), 2), c(0,1), col='red', lty=1, lwd=2)
lines(rep(pA*nV*(1-e.rep), 2), c(0,1), col='red', lty=2, lwd=2)  
\end{verbatim}
A number of hundred thousand vaccinated individuals
has been used, with
an {\em absolutely hypothetical} value of assault probability
of 1\,\%. 
The script can also be used to simulate the effect of a precise
value of $\epsilon$, thus exactly corresponding to the efficacy,
just setting its standard deviation to a very small value.
\begin{figure}
  \begin{center}
    \epsfig{file=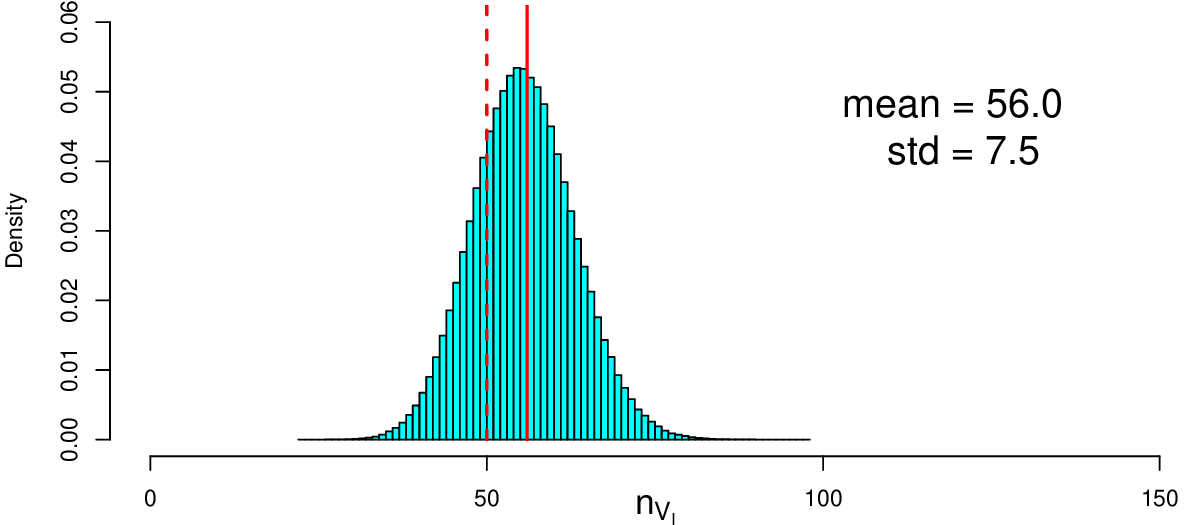,clip=,width=0.73\linewidth}\\
    \epsfig{file=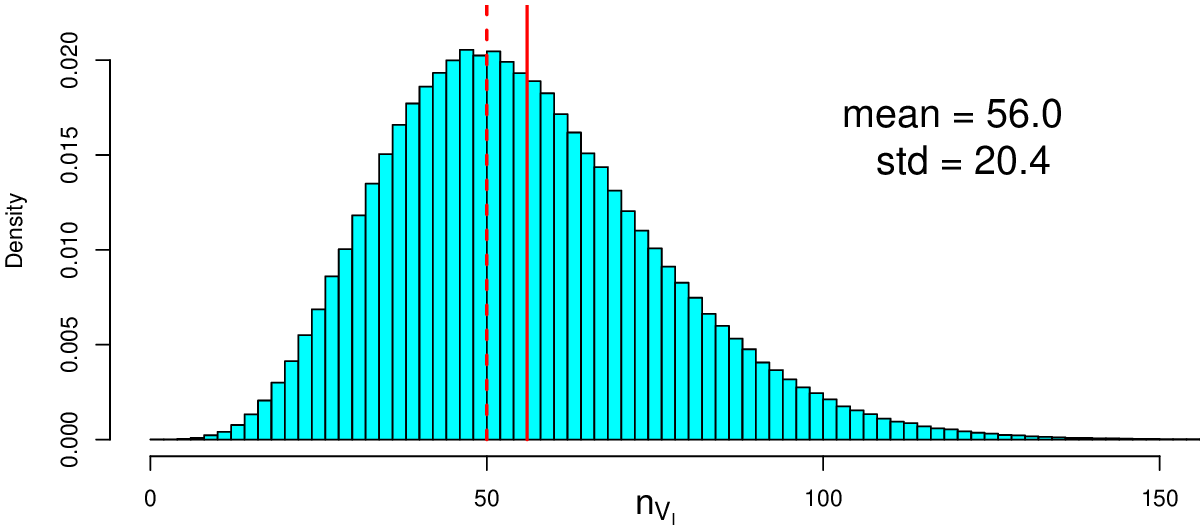,clip=,width=0.73\linewidth}\\
     \epsfig{file=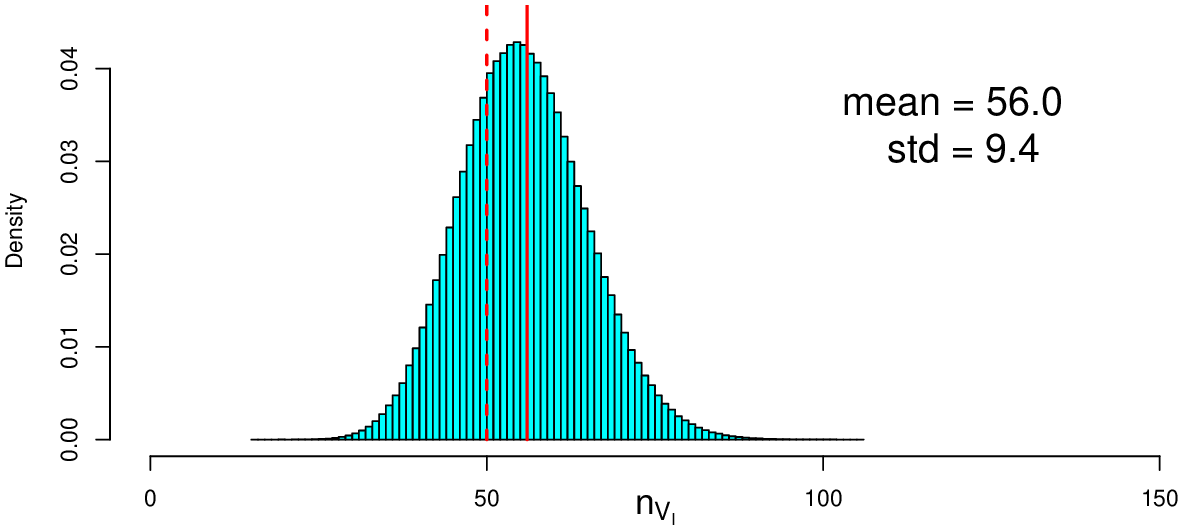,clip=,width=0.73\linewidth}\\
    \epsfig{file=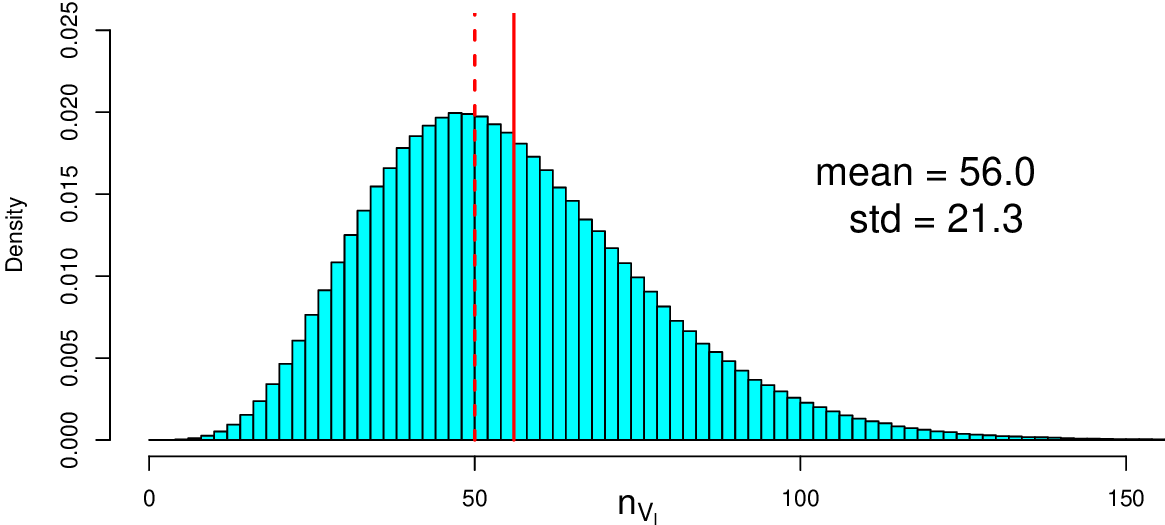,clip=,width=0.73\linewidth} 
       \\  \mbox{} \vspace{-1.3cm} \mbox{}
  \end{center}  
  \caption{\small \sf
    Distribution of the predicted number of vaccinated infectees, subject
    to (top-down): exact values of $\epsilon$ and $p'_A$;
    uncertain $\epsilon$; uncertain $p'_A$;
    uncertain $\epsilon$ and $p'_A$ (see text for details).
  }
   \label{fig:predicted_VI} 
\end{figure}
The result of this idealized situation,
using e.g. $\epsilon = 0.944$, i.e.
the mean resulting from the MCMC analysis
of Pfizer data,\footnote{There is nothing special with this choice,
  and what follows is a little more than an exercise, strongly dependent
  on the assumption on $p'_A$.
  }
is shown in the top histogram of Fig.~\ref{fig:predicted_VI}.
In this idealized case the distribution of
$n_{V_I}'$ is simply a binomial, i.e.
\begin{eqnarray*}
  n_{V_I}' &\sim & \mbox{Binom}(n_{V}',\, p_{ov})\,, 
\end{eqnarray*}
with the {\em overall} probability $p_{ov}$
being equal to $p_A'\cdot(1\!-\!\epsilon)$, that is
$p_{ov} = 0.00056$ for the numbers reported in the script.
Expected value and standard deviation are then 
56.0 and 7.5, in perfect agreement with the Monte Carlo result
shown in the upper panel of Fig.~\ref{fig:predicted_VI}.

The effect of the uncertainty about $\epsilon$ is shown in the
second (top-down) histogram of the same figure. As we can see,
the distribution becomes remarkably wider and more asymmetric,
with a right-hand skewness, effect of the left-hand skewness of $f(\epsilon)$.
We see then, in the third histogram, the effect
of a hypothetical uncertainty about $p_A'$, modeled here
with a standard deviation of $\sigma(p'_A)=0.1\times p'_A$  (but this has to be understood
really as an {\em exercise} done only to have an idea of the effect,
because a reasonable uncertainty could indeed be {\em much larger}).
Finally, including both sources of uncertainty, we get
the histogram and the numbers at the bottom of the figure.
Vertical lines show the predicted values for $n_{V_I}'$
by using the MCMC mean value (solid line) and using the
modal value (dashed lines).  

As a further step, following  Ref.~\cite{sampling}
(see in particular Secs.~5.2.1 and 5.3.1 there),
let us try to get approximated formulae for the expected value
and the standard deviation of $n'_{V_I}$. The idea, we shortly remind, 
is to start with the expected value and variance evaluated
for the expected values of  $\epsilon$ and $p'_A$, and then
make a `propagation of uncertainty' by linearization
(as if  $\epsilon$ and $p'_A$ were `systematics').
Here are the resulting formulae
\begin{eqnarray}
\mbox{E}(n'_{V_I}) &\approx&
      n'_V \cdot \mbox{E}(p'_A)\cdot [1-\mbox{E}(\epsilon)] \label{eq:mean_approx} \\  
\sigma^2(n'_{V_I}) & \approx & 
\sigma^2\left[n'_{V_I}\,|\,\,p_A'=\mbox{E}(p'_A),
                         \epsilon=\mbox{E}(\epsilon)\right] \nonumber \\
&& +\, \left(n'_V\cdot[1-\mbox{E}(\epsilon)]\right)^2 \cdot \sigma^2(p'_A) \, +\,
\left(n'_V\cdot \mbox{E}(p'_A)\right)^2 \cdot \sigma^2(\epsilon) \nonumber \\
&\approx&  n'_V \cdot  \mbox{E}(p'_A)\cdot\left[1-\mbox{E}(\epsilon)\right]\cdot
\left[1- \mbox{E}(p'_A)\cdot[1-\mbox{E}(\epsilon)]\right]  \nonumber\\
&& +\, {n'}_V^2\cdot\left[1-\mbox{E}(\epsilon)\right]^2 \cdot \sigma^2(p'_A) \, +\,
{n'}_V^2\cdot\mbox{E}^2(p'_A) \cdot \sigma^2(\epsilon)\,, \label{eq:std_approx}
\end{eqnarray}
which we have checked to be in excellent agreement
with the results from direct Monte Carlo.\footnote{
  For example, here is the R code to be added in the above script,
  immediately after the assignment `{\tt pA <- 0.01}',
  in order to implement Eqs.~(\ref{eq:mean_approx})\,-\,(\ref{eq:std_approx}):
\begin{verbatim}
spA <- 0.1 * pA
est.nvI <- nV * pA * (1-mu)
est.sigma.nvI <- sqrt( nV * pA * (1-mu) * (1 - pA * (1-mu)) +
                      (nV*(1-mu))^2 * spA^2 + (nV*pA)^2 * sigma^2 )
cat(sprintf("Approximated nvI: mean+-sigma: %.1f +- %.1f\n",est.nvI,est.sigma.nvI))
\end{verbatim}  
} 

\subsection{More on the probability that a vaccinated person gets
  shielded from the virus -- a Monte Carlo approach}
Section \ref{sec:mode_mean} has been devoted to explain the reason
why the proper number to be reported as `efficacy',
meant as the probability that a vaccinated person
is shielded from the virus, is the mean of the
pdf of the model parameter $\epsilon$. Having talked in this
section about predicting the number of infects,
we can use the same extended model of
Fig.~\ref{fig:numero_atteso_vaccinati_infetti}
in order to check the outcome of that reasoning.
It is in fact enough to
set $n'_V=1$ and $p'_A=1$ and analyze the result of the
MCMC. Indeed, $n'_{V_A}$ will be identically 1,
in the sense that the person will be `assaulted' with certainty,
but the output $n'_{V_I}$ can have now only two possible values, 0
(person not infected) or 1 (person infected).
If $\epsilon$ were exactly known, and let us indicate it
by $\epsilon_K$, the probability of $n'_{V_I}=0$ would be  
$\epsilon_K$ and that of  $n'_{V_I}=1$ would be $1-\epsilon_K$.
A simple direct Monte Carlo would then produce a fraction of
occurrences of $n'_{V_I}=0$ around $\epsilon$. If, instead, $\epsilon$
is unknown, then the values used in the bottom-left side
of the network will be those occurring in the MCMC chain.
Therefore, we reobtain the same result seen
in  Sec.\ref{sec:mode_mean}, i.e. that the relative frequency
of the occurrence of $n'_{V_I}=0$ will be equal to the
mean of $\epsilon$ in the chain.

This MCMC strategy offers a further argument,
to which some practitioners might be more reactive,
in support of the thesis that the number to be reported as `efficacy'
should be the average of the probability distribution
of $\epsilon$, rather than other possible summaries of the distribution.

\section{Efficacy against disease severity}\label{sec:severity}
A last point we wish to address in this paper is related to
the efficacy of the vaccines against disease severity,
based on the data reported by Moderna and Pfizer
(see Tab.~\ref{tab:data}): 
in the first case 30 people got a `severe form' out of 185
infectees in the control group; none of the severe cases occurred in the group of 11 vaccinated infectees;
in the second the corresponding numbers are 9 in 162
and 1 in 8.

In order to analyze this further pieces of information we 
can simply extend the Bayesian network of
Fig.~\ref{fig:BayesianNet} adding four nodes
(see Fig.~\ref{fig:BayesianNet1}):
{\small 
\begin{figure}
  \begin{center}
    \epsfig{file=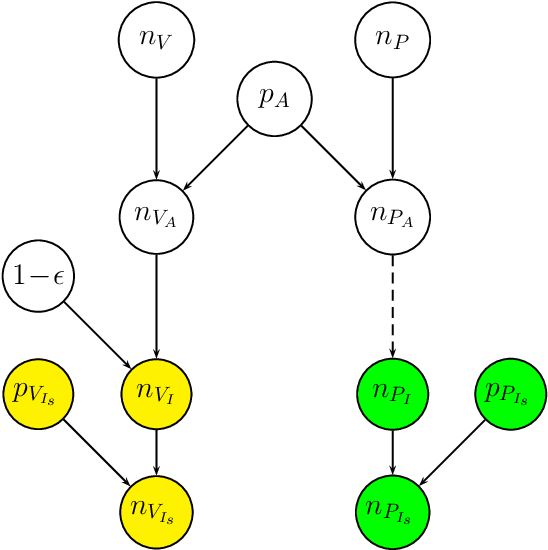,clip=,width=0.6\linewidth}
          \\  \mbox{} \vspace{-1.0cm} \mbox{}
  \end{center}
  \caption{\small \sf Extended Bayesian network of the vaccine vs placebo
    experiment (see text). 
  }
  \label{fig:BayesianNet1}
\end{figure}
}
 $n_{V_{I_s}}$ and $n_{P_{I_s}}$ represent
the number of infected individuals that got
the disease in a severe form, while  $p_{V_{I_s}}$
and $p_{P_{I_s}}$ represent the corresponding probability of
developing severe diseases in either group. 
Again we can use the binomial distributions: 
\begin{eqnarray}
  n_{V_{I_s}} &\sim & \mbox{Binom}(n_{V_I},\, p_{V_{I_s}})  \label{eq:distribution}\\
  n_{P_{I_s}} &\sim & \mbox{Binom}(n_{P_I},\, p_{P_{I_s}})  \label{eq:distribution1}\,,   
\end{eqnarray} 
and the following JAGS model would result:
\begin{verbatim}
    model {
      nP.I  ~ dbin(pA, nP)           #  1.          
      nV.A  ~ dbin(pA, nV)           #  2.
      pA    ~ dbeta(1,1)             #  3. 
      nV.I  ~ dbin(ffe, nV.A)        #  4.  [ ffe = 1 - eff ]
      ffe   ~ dbeta(1,1)             #  5.
      eff   <- 1 - ffe               #  6. 
      pS_P  ~ dbeta(1,1)             #  7.
      pS_V  ~ dbeta(1,1)             #  8.  
      nS.V  ~ dbin(pS_V,nV.I)        #  9.
      nS.P  ~ dbin(pS_P,nP.I)        # 10.
    }
\end{verbatim}
However, looking at the Bayesian network of Fig.~\ref{fig:BayesianNet1},
it is clear that, {\em being $n_{V_I}$ and  $n_{{P_I}}$ observed nodes},
i.e. $n_{V_I}$ and  $n_{{P_I}}$ are just data, the bottom nodes involving
$(n_{V_{I_s}}, n_{V_I}, p_{V_{I_s}})$ and $(n_{P_{I_s}}, n_{P_I}, p_{P_{I_s}})$
get `separated' from the rest of the network. In other words
there is {\em no flow
  of evidence} from $(n_{V_{I_s}}, p_{V_{I_s}})$,
or from  $(n_{P_{I_s}}, p_{P_{I_s}})$, to the rest of the network.
Therefore the problem has a rather simple solution.
In particular, using uniform priors for $p_{V_{I_s}}$
and  $p_{P_{I_s}}$, we get
\begin{eqnarray*}
  p_{V_{I_s}} &\sim & \mbox{Beta}(n_{V_{I_s}}\!+\! 1, n_{V_I}\!-\!n_{V_{I_s}}\!+1)  \\
  p_{P_{I_s}} &\sim & \mbox{Beta}(n_{P_{I_s}}\!+\! 1, n_{P_I}\!-\!n_{P_{I_s}}\!+1)  \,.   
\end{eqnarray*}
Nevertheless, for didactic purposes, we report in Fig.~\ref{fig:severity}
\begin{figure}
  \begin{center}
    \epsfig{file=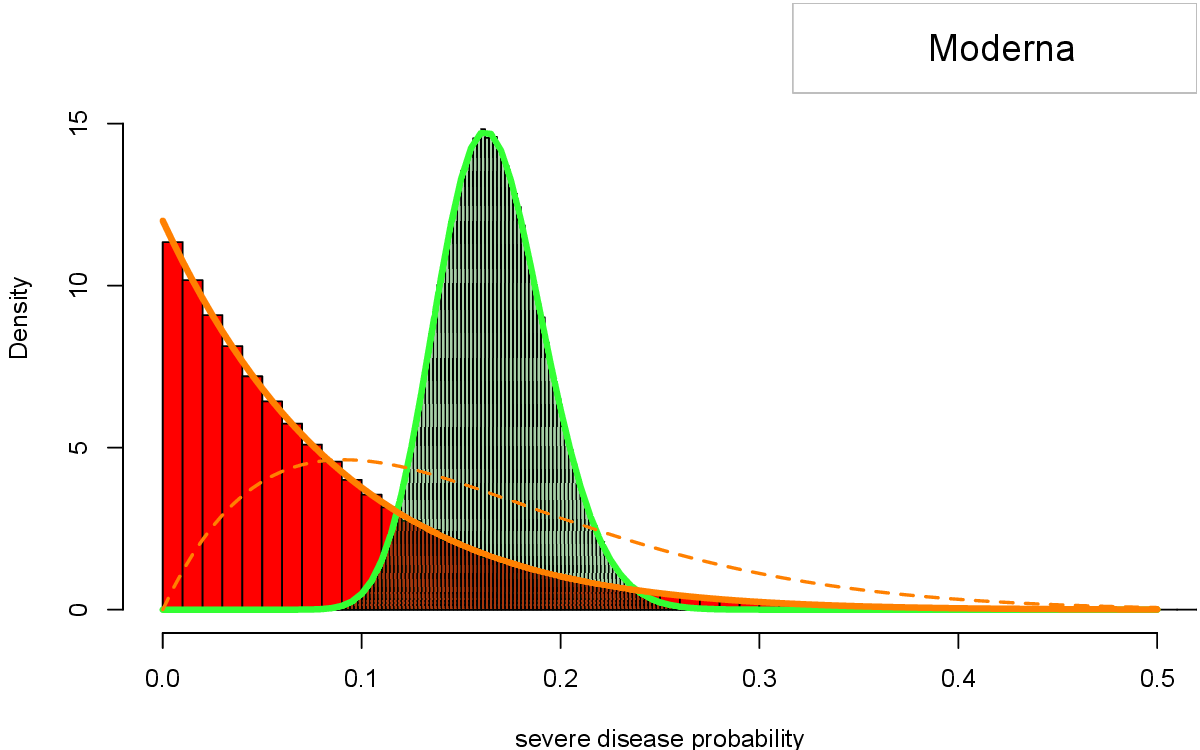,clip=,width=0.9\linewidth} \\
    \mbox{}  \\ \mbox{} \\
    \epsfig{file=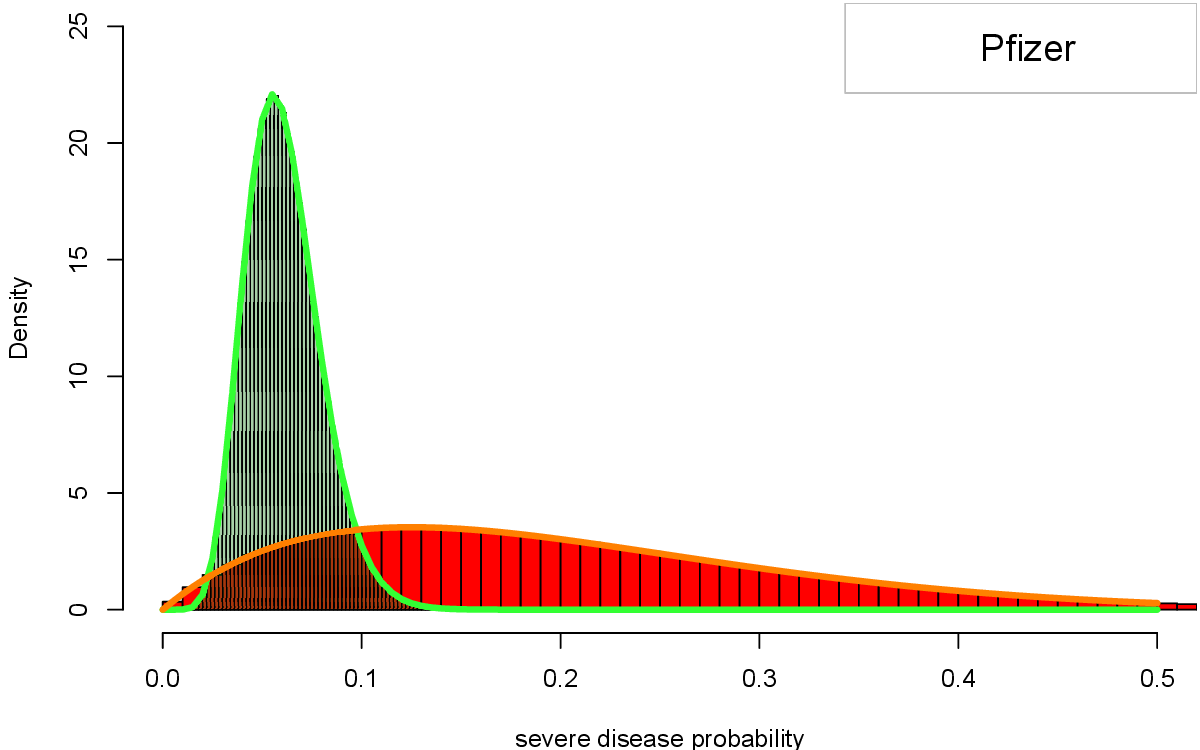,clip=,width=0.9\linewidth}
  \end{center}  
      \caption{\small \sf Distribution of the probability of getting
        a severe form of Covid-19 (see text).
      }
  \label{fig:severity}

  \end{figure}
the histograms of the MCMC results with superimposed the Beta pdf's
(solid lines --
we shall come back later to meaning of
the dashed line added in the case of Moderna).

As far as the control groups are concerned
(green, narrower histograms and curves in Fig.~\ref{fig:severity}),
the results from Moderna and Pfizer data are quite different. 
In both cases we get rather narrow distributions, as expected
from the rather large numbers involved (and therefore the central
values are close to the proportion of severe cases with respect
to the total number). But they differ substantially
and, using mean and standard deviation to summarize them,
we get  
\begin{eqnarray*}
  \mbox{Moderna:} & &   p_{P_{I_s}}= 0.170 \pm 0.028 \\   
   \mbox{Pfizer:} & &  p_{P_{I_s}}=0.055 \pm 0.018\,,
\end{eqnarray*}
differing by $0.115\pm 0.033$. In particular,
what results by analyzing  Moderna data
is in good agreement with the estimates by the  WHO, 
which states that ``{\em approximately 10-15\% of cases progress to
severe disease, and about 5\% become critically ill}''~\cite{WHOsev}.
Obviously, we are not in the position to make any
statement about the reason of the disagreement,
that could be due to the different populations
on which the trials have been performed.
But, if this were the case, one could have some doubts about the validity
of comparing the different sets of data. We can only leave the question to
epidemiology experts.

Passing to the vaccine groups (red, broader histograms and curves
in Fig.~\ref{fig:severity}),
the crude summaries in terms of  mean and standard deviation give
\begin{eqnarray*}
  \mbox{Moderna:} & &   p_{V_{I_s}}=0.077 \pm 0.071 \\
   \mbox{Pfizer:} & &  p_{V_{I_s}}=0.200 \pm 0.121\,.
\end{eqnarray*}
If we consider just the mean values, it seems that there is
a very large difference between the performances of the two vaccines.
However their difference is hardly significant, being their difference
$-0.12 \pm 0.14$. Also the fact that the two curves 
{\em looks substantially different} should not impress a statistics
expert if she knew from which number they have been derived.
The main difference is due to the fact that Moderna has zero severe disease
against one of Pfizer. If they had had one too, which would not be surprising 
in a hypothetical `second trial', the curve would be not dissimilar
from the Pfizer one (dashed line in Fig.~\ref{fig:severity}),
vanishing at zero and yielding  $p_{V_{I_s}}=0.154 \pm 0.096$
-- practically the same, an experienced physicist would say. 

It is quite evident that it is not possible to draw  general conclusions 
on the efficacy of the generic vaccine on softening the impact of the disease.
But the real point we wish to highlight, given the spread
of distributions, is that we do not have enough data for drawing sound conclusion.
For this reason we wish to point out that, even for this aspect,
press releasing a 100\% efficacy and not dealing with
the unavoidable uncertainties and their impact when applied
to decision making is quite misleading. Figure \ref{fig:severity}
indeed shows that the probability of becoming severely ill in
the vaccine group is definitively low but, quite obviously,
not zero and with a relevant overlap with the distribution
evaluated for the control group.

\section{Summing up} \label{sec:Summing}
In this paper, spurred by the press releases first
and then by the published results about the performance
of the candidate vaccines, we have reanalyzed the published 
data with the help
of a Bayesian network processed with MCMC methods.
The aim was that of obtaining, for each data set,
the pdf of the model variable $\epsilon$,
whose meaning is the following:
if we {\em assume}
for it an {\em exact} value, then the probability that
a `virus assaulted individual' gets infected  
{\em would be exactly} $1\!-\!\epsilon$.

Our results are in excellent agreement with the published
ones, if the latter are properly interpreted. In fact, it turns out that
they coincide with the modes of the respective
pdf $f(\epsilon)$,
although we maintain that if a single number has to be provided,
especially to the media, it should be 
the mean of the distribution. 
This has, in fact, the meaning of the probability
that a vaccinated person will be shielded by the virus,
taking into account the unavoidable uncertainty on
$\epsilon$, fully described by $f(\epsilon)$. And this
is what really matters to define the efficacy of a vaccine.
Willing however to reduce the result of our analysis to a single
number to be compared
with the released ones, we get respectively
and with reasonable rounding, 93\%, 94\%, 94\%, 86\% and 60\% for 
Moderna-1, Moderna-2, Pfizer,
AstraZeneca (LDSD) and AstraZeneca (SDSD),
versus 94.5\%, 94.1\%, 95.0\%, 90.0\% and 62.1\%
of Tab.~\ref{tab:vaccine_efficacy_published}.
Therefore, as far as these numbers are concerned, there is
then a {\em substantial agreement} of the outcome of
our analysis with the published results, simply because
when a probability distribution is unimodal and
rather symmetrical then  mode and mean tend to coincide.
Therefore, with respect to the main results,
our contribution to this point is mainly methodological.
The probability theory based result is, instead,
at odds with Moderna 100\% claimed efficacy against severe disease,
for which a more sound 92\% should be quoted.

In order to summarize more effectively the probability distribution
of $\epsilon$ with just a couple of numbers,
our preference goes to mean and standard deviation, 
although we also report the bounds of the
central 95\% {\em credible interval}.
This interval is, once more, in excellent
agreement not only with the Pfizer result, which
has also published an interval having exactly the same meaning,
but also with the uncertainty intervals of the other companies,
although they provide {\em confidence intervals},
which, strictly speaking, do not have the same meaning
of the credible intervals. This is not a surprise to us.
We are in fact aware that in many practical cases not only
frequentistic point estimates are equivalent to the mode of the posterior
distribution of the model parameter, {\em if} a uniform prior was
used in a Bayesian analysis based on the same data,
but also `95\% confidence intervals'
{\em tend to} be, numerically, equal to the 95\% probable
intervals.\footnote{But this is not a general rule,
  as discussed in detail in Ref.~\cite{BR}.} 

This takes us to the question of the priors. As just reminded,
a uniform 
prior over $\epsilon$ has been used in our analysis.
But, clearly, not because we believe that the efficacy of 
a vaccine that has reached the Phase-3 trial
has the same chance to be close to zero or to one.
Instead, a flat prior can be considered a convenient practical choice,
if the inference is dominated by the data, as it
is often the case.
Moreover, the advantage of a uniform prior in
parametric inference is that the effect of an informative prior
reflecting the opinion of experts
can be taken into account at a later time.
This `posterior use of priors' might sound
paradoxical, but it is important to remind that in Bayesian
inference `prior' does not indicate time order but rather `based on
the status of knowledge without taking into account the new piece of information'
provided by the data entering the specific analysis. 
Having, in fact, prior and likelihood symmetric and peer roles
in the Bayes' rule,
an expert can use her prior to `reshape' the posterior pdf
resulting from data analysis, if a flat prior was used,
without having to ask to repeat the analysis (see Ref.~\cite{sampling}
for details).

This reshaping becomes particularly simple if the prior is modeled
by a convenient, rather flexible probability distribution such as
the Beta. In fact, as we have seen, the pdf of $\epsilon$ starting from
a flat prior tends to resemble a Beta. The same is then true if also
the prior is modeled by a Beta (this is related to the well known fact
of the Beta being the {\em conjugate prior} of a binomial distribution,
even though our model is not just a simple binomial). These observations
are particularly interesting because they lead to
Eqs.~(\ref{eq:update_rp})\,-\,(\ref{eq:update_sp}), which,
together with Eqs.~(\ref{eq:r_from_E-sigma})\,-\,(\ref{eq:s_from_E-sigma}),
allow to take easily into account the expert priors.
In fact, if the priors are
rather vague, $r_0$ and $s_0$ appearing in
Eqs.~(\ref{eq:update_rp})\,-\,(\ref{eq:update_sp}) are quite
small (although larger than one, since $\epsilon=0$ and $\epsilon=1$
are {\em \`a priori} reasonably ruled out) and, in particular,
smaller that $r_F$ and $s_F$.\footnote{Just to have
an idea of the numbers we are dealing with, the values
of $\{r_F,\,s_F\!\}$ resulting from our analysis
are equal to  $\{73,\,5.3\}$, $\{156,\,11\}$, $\{137,\,8.1\}$,
$\{17,\,2.8\}$ and $\{17,\,11\}$, respectively for  
Moderna-1, Moderna-2, Pfizer,
AstraZeneca (LDSD) and AstraZeneca (SDSD).
}
If, instead, the expert has
a strong opinion about the possible values of $\epsilon$, then
$r_0$ and $s_0$ will play a role in {\em her} posterior,
and in that of {\em her community}, if its members trust
her.\footnote{Remember that Science and its popularization
  is based on a long chain of rational beliefs~\cite{BR}. Think for
  example to the reasons \underline{you} believe in gravitational waves,
  {\em provided that} you really
  believe that they could exist and  that they have finally
  being detected on Earth starting from 2015
  -- our {\em trusted source} ensures us that 67 of them
  have been `observed' so far \cite{Pia}. 
} 

Coming back to the way to summarize $f(\epsilon)$, 
our preference goes to its mean and standard deviation.
The mean because, as reminded above, has the meaning of efficacy
for vaccine treated people not having been involved in the trial,
if all possible values of $\epsilon$ are taken into account.
The standard deviation because it is mostly convenient,
together with the mean, to make use of the result
of the inference in further considerations and
in `propagation of uncertainties',
thanks to general probability rules.

We have just reminded the utility of mean and standard deviation in order to
re-obtain $f(\epsilon)$, under the hypothesis that it is
{\em almost} a Beta distribution, making use of
Eqs.~(\ref{eq:r_from_E-sigma})\,-\,(\ref{eq:s_from_E-sigma}).
The application related to `propagation of uncertainties'
that we have seen in the paper has to do with predicting
the number of individual that will get infected in a group
that it is going to be vaccinated.
This is a problem in probabilistic forecasting and the
number of interest is uncertain for several reasons.
There is, unavoidably, the uncertainty deriving from the inherent
binomial distribution, having {\em assumed} an assault probability
$p_A'$ in the new population.
But also the uncertainties on the values of $p_A'$
and $\epsilon$ play a role, that can even be dominant
with respect to the `statistical' effect of the binomial.

Now, the probability distribution of the number of
vaccinated infectees can be evaluated extending our
basic Bayesian network, as we have done here.
But we have also stressed the importance of having
approximated expressions, based on linearization,
for its expected value and standard deviation.
And such expressions, thus obtained considering $\epsilon$ 
and  $p_A'$ as `systematic'~\cite{BR}, depend then on their 
mean and standard deviation. For
example the contribution to $\sigma(n_{V_I}')$
due to the uncertain $\epsilon$, and then 
to be added `in quadrature' to the other sources of uncertainty,
is given by ${n'}_V\!\cdot\!\mbox{E}(p'_A)\! \cdot\! \sigma(\epsilon)$.
This gives at a glance the contribution to the global uncertainty
without having to run a Monte Carlo.\footnote{Note also
that what really enters in Eqs.~(\ref{eq:mean_approx})\,-\,(\ref{eq:std_approx})
is $1\!-\!\epsilon$ whose relative uncertainty is around
30\% even in the best cases of Moderna-2 and Pfizer $[$it reaches
54\% for AstraZeneca (LDSD), becoming `only' 22\%
for AstraZeneca (SDSD), characterized however by a large value
of  $1\!-\!\epsilon$\,$]$.
}

Finally, a comment on how to possible reduce
$\sigma(\epsilon)$ is in order. In fact, the relative uncertainty on
$\epsilon$ depends on the small number of vaccinated infectees.
This suggests that the quality of its `measurement'
could be improved, keeping constant the total numbers of
individuals entering the trial,
if the size of the placebo group is reduced.
We have checked by simulation that reducing it by 2/3,
thus having about a factor of five between the two groups, 
$\sigma(\epsilon)$ is expected to be reduced by about 20\%.
Not much indeed, but this different sharing of individuals
in the two groups would have the advantage
of increasing the chance of detecting side effect of the vaccine,
basically at the same cost.

\section{Conclusion} \label{sec:conclusions}
A re-analysis of the 
2020 data on vaccine trials by Pfizer, Moderna and
AstraZeneca has been performed with didactic intent and 
focusing on uncertainties and on what most effectively should be reported
as outcome. With this respect, we appreciate once more the role provided
by Bayesian networks in clarifying the meaning of each
variable entering the model and its relation with the others.
In particular, we make a distinction between the
`efficacy variable' $\epsilon$ and the {\em efficacy} to be reported
to the scientific community and to the general public as
{\em probability that a newly vaccinated person will shielded from Covid-19}.
The uncertain values of $\epsilon$ are characterized by the pdf
$f(\epsilon)$, obtained in this work by MCMC, and
whose {\em mean value} has exactly the meaning of {\em efficacy},
in analogy to Laplace's rule of succession to quantify the probability
of a future event.

We have stressed not only the importance of providing
the most complete information of $f(\epsilon)$, whose graphical
representation provides better than many words its uncertainty,
but also that of summarizing the results (if only two numbers
have to be chosen) with mean and standard
deviation of the distribution. In fact, although 
the probability distribution of $\epsilon$ is what is
really needed to the development of predictive what-if scenarios,
mean and standard deviation are the most useful quantities
to be used for further approximated evaluations. 

With regards to the comparison with the published result, the efficacy
values obtained in this analysis as mean of the probability distribution
of $\epsilon$ are in good agreement with them
for the reasons just reminded in the previous section, with
the exception of the 100\% claim that speaks for itself.
This agreement plays an important role in validating the causal
model on which the present analysis is based,
indeed very simple (but not simplistic!), that
can be used by students and researchers to repeat the analysis --
for this reason pieces of programming code have also been provided.

 \mbox{} \\
 \noindent
 {\bf \large Note added}\\
 We have learned in the meanwhile that the strategy of
 choosing the vaccine sample much larger than the
 placebo one, which we suggest at the end of the previous
 section, has been used in real life trials by
 Sputnik \cite{Sputnik} and AstraZeneca \cite{AstraZeneca_asym},
 which used vaccine vs placebo sample ratios of
 3:1 and 2:1, respectively.



{\footnotesize
  \begin{thebibliography} {ref99}

\bibitem{Pfizer_09nov}
Pfizer Inc., Press Release, November 9, 2020,\\
\url{https://www.pfizer.com/news/press-release/press-release-detail/pfizer-and-biontech-announce-vaccine-candidate-against}.

\bibitem{Pfizer_18nov}
Pfizer Inc., Press Release, November 18, 2020,\\
  \url{https://www.pfizer.com/news/press-release/press-release-detail/pfizer-and-biontech-conclude-phase-3-study-covid-19-vaccine}.
  
\bibitem{Pfizer_NEJM}
P. Fernando \textit{et al.}
for the C4591001 Clinical Trial Group (Pfizer),
{\em Safety and Efficacy of the BNT162b2 mRNA Covid-19 Vaccine}, The New England Journal of Medicine, 
\url{https://www.nejm.org/doi/pdf/10.1056/NEJMoa2034577?articleTools=true}.

\bibitem{Moderna_16nov}
Moderna Inc., Press Release, November 16, 2020,\\
  \url{https://investors.modernatx.com/news-releases/news-release-details/modernas-covid-19-vaccine-candidate-meets-its-primary-efficacy}.
  
\bibitem{Moderna_30nov}
Moderna Inc., Press Release, November 30, 2020,\\
  \url{https://investors.modernatx.com/news-releases/news-release-details/moderna-announces-primary-efficacy-analysis-phase-3-cove-study}.
  
\bibitem{Moderna_NEJM}
L.R. Baden \textit{et al.} {\em Efficacy and Safety of the mRNA-1273 SARS-CoV-2 Vaccine}, The New England Journal of Medicine, \\ 
\url{https://www.nejm.org/doi/10.1056/NEJMoa2035389}.
  
\bibitem{AZ1}
  AstraZeneca PLC, News Release, 23 November 2020,\\
\url{https://www.astrazeneca.com/content/dam/az/media-centre-docs/press-releases/2020/AZD1222-HLR-RNS.pdf}.
  
\bibitem{AstraZeneca}
M. Voysey {\em et al.} (AstraZeneca)
{\em Safety and efficacy of the ChAdOx1 nCoV-19
vaccine (AZD1222) against SARS-CoV-2:
an interim analysis of four randomised controlled
trials in Brazil, South Africa, and the UK},
December 8, 2020,\\
\url{https://doi.org/10.1016/S0140-6736(20)32661-1}.

\bibitem{vaccini}
G. D'Agostini and A. Esposito, {\em Inferring vaccine efficacies
and their uncertainties. 
A simple model implemented in JAGS/rjags}, 19 November 2020,\\
\url{https://www.roma1.infn.it/~dagos/prob+stat.html#vaccini}.

\bibitem{Fauci_yt_16nov}
A. Fauci at NBC News,  November\! 16,\! 2020,  
\url{https://youtu.be/8CG8aI4XCGw?t=32}.

\bibitem{ISO}  
International Organization for Standardization (ISO), 
{\it Guide to the expression of uncertainty in measurement}, 
Geneva, Switzerland, 1993, \\
\url{https://www.bipm.org/en/publications/guides/gum.html}.

\bibitem{WavesSigmas}
G. D'Agostini, {\em The Waves and the Sigmas
    (To Say Nothing of the 750 GeV Mirage)},  
\href{https://arxiv.org/abs/1609.01668}{arXiv:1609.01668 [physics.data-an]}.

\bibitem{BR}
G. D'Agostini, {\em Bayesian Reasoning in Data Analysis.
A critical Introduction}, World Scientific, 2003.

\bibitem{JAGS}
  M. Plummer, {\em JAGS: A Program for Analysis
    of Bayesian Graphical Models Using Gibbs Sampling},
  Proceedings of the 3rd International Workshop on Distributed
  Statistical Computing (DSC 2003), March 20–22, Vienna, Austria.
  ISSN 1609-395X, 
\url{http://mcmc-jags.sourceforge.net/}\,.

\bibitem{Moderna_100_media}
See, e.g.,\\    
  J. Cohen, Science, {\em ‘Absolutely remarkable’:
    No one who got Moderna’s 
    vaccine in trial developed severe COVID-19}, November 30, 2020, \\
  \url{https://www.sciencemag.org/news/2020/11/absolutely-remarkable-no-one-who-got-modernas-vaccine-trial-developed-severe-covid-19}
  (https://www.sciencemag.org/news/2020/11/absolutely-remarkable-no-one\\
  -who-got-modernas-vaccine-trial-developed-severe-covid-19).\\
D. Grady,  New York Times, {\em Moderna Applies for Emergency F.D.A. Approval for Its Coronavirus Vaccine},  November 30 2020,\\
 \url{https://www.nytimes.com/2020/11/30/health/covid-vaccine-moderna.html}\,.\\
Repubblica, {\em Vaccino, Moderna chiede l'autorizzazione all'uso di
   emergenza alla Fda e all'Ema. L'azienda americana annuncia i
   risultati della fase 3: ``Efficacia pari al 94,1\% e fino al 100\% nei casi gravi''},
 November 30, 2020, \\
 \url{https://www.repubblica.it/cronaca/2020/11/30/news/vaccino_i_dati_di_moderna_efficace_al_94_1_e_fino_al_100_nei_casi_gravi_-276437984/}\,.\\
 E. Cohen, CNN, {\em Moderna applies for FDA authorization for its Covid-19 vaccine}, December 1, 2020,
  \url{https://edition.cnn.com/2020/11/30/health/moderna-vaccine-fda-eua-application/index.html}\,. 
  
\bibitem{sampling}
  G. D'Agostini and A. Esposito, {\em Checking individuals
    and sampling populations with imperfect tests},
  \href{https://arxiv.org/abs/2009.04843}{arXiv:2009.04843 [q-bio.PE]}.
  

\bibitem{R}
  R Core Team (2018), {\em R: A language and environment for statistical
  computing}. R Foundation for Statistical Computing, Vienna, Austria,\\
  \url{https://www.R-project.org/}\,.
  
\bibitem{rjags}
  M. Plummer, {\em rjags: Bayesian Graphical Models using MCMC}.\\
  R  package version 4-10,
  \url{https://CRAN.R-project.org/package=rjags}\,.
  
\bibitem{Gavi}
{\footnotesize  
  \url{https://www.gavi.org/vaccineswork/what-difference-between-efficacy-and-effectiveness} \\
\mbox{}\vspace{-0.5cm}  
}
  
\bibitem{Bayes}
Bayes, Thomas and Price, Richard  {\em An Essay towards
solving a Problem in the Doctrine of Chance. By the late
Rev. Mr. Bayes, communicated by Mr. Price,
in a letter to John Canton, A.M.F.R.S.},
Philosophical Transactions of the Royal Society of London. 53: 370–418,
(1763),
\url{https://doi.org/10.1098%2Frstl.1763.0053}\,.

\bibitem{Laplace_PC}
  P.S. Laplace,
  {\em Mémoire sur la probabilité des causes par les événements''},
  Mémoire de l'Académie royale des Sciences de Paris (Savants
étrangers),
  Tome VI, p. 621, 1774,
  \url{https://gallica.bnf.fr/ark:/12148/bpt6k77596b/f32}\,.

      
\bibitem{Laplace2}
P.S. Laplace, {\it Essai philosophique sur les probabilités}, 1814,\\
\url{http://books.google.it/books?id=JrEWAAAAQAAJ}.
(English quotes from the classical translation by F.W. Truscott and F.L. Emory:
  \url{https://bayes.wustl.edu/Manual/laplace_A_philosophical_essay_on_probabilities.pdf}.)

\bibitem{WHOsev}
  World Health Organization,\,
  \url{https://www.who.int/docs/default-source/coronaviruse/risk-comms-updates/update-36-long-term-symptoms.pdf?sfvrsn=5d3789a6_2#:~:text=%E2%80%A2%20Most%20people%20with%20COVID,have%20lasting%20health%20effects.}
 
\bibitem{Pia}
LIGO-Virgo Cumulative Event/Candidate Rate Plot O1-O3,\\
\url{https://dcc.ligo.org/LIGO-G1901322/public}.

\bibitem{Sputnik}
Denis Y. Logunov \textit{et al.} {\em Safety and efficacy of an rAd26 and rAd5 vector-based heterologous prime-boost COVID-19 vaccine: an interim analysis of a randomised controlled phase 3 trial in Russia}, February 2, 2021,\\
\url{https://doi.org/10.1016/S0140-6736(21)00234-8}.

\bibitem{AstraZeneca_asym}
  AstraZeneca, {\em AZD1222 US Phase III trial met primary
    efficacy endpoint in preventing COVID-19 at interim analysis},
  March 22, 2021,\\
\url{https://www.astrazeneca.com/media-centre/press-releases/2021/astrazeneca-us-vaccine-trial-met-primary-endpoint.html}.  

\end{thebibliography}
}
\end{document}